%% file: main.tex
\documentclass[table,xcdraw]{fairmeta} 
\usepackage{lipsum}
\usepackage{bbm}
\usepackage{float}
\usepackage{amssymb}
\immediate\write18{bibtex \jobname}

\usepackage{tabularx}

\usepackage{times}
\usepackage{latexsym}
\usepackage{booktabs}
\usepackage[T1]{fontenc}

\usepackage[utf8]{inputenc}

\usepackage{microtype}

\usepackage{inconsolata}

\usepackage{multirow}

\usepackage{graphicx}
\usepackage{tikz} 

\usepackage{colortbl}

\usepackage{xspace}
\usepackage{soul}
\usepackage{textcomp}

\usepackage{adjustbox}

\title{CyberSOCEval: Benchmarking LLMs Capabilities for Malware Analysis and Threat Intelligence Reasoning}

\author{Lauren Deason*†}

\author{Adam Bali†}
\author{Ciprian Bejean‡}
\author{Diana Bolocan*‡}
\author{James Crnkovich*†}
\author{Ioana Croitoru*‡}
\author{Krishna Durai†}
\author{Chase Midler‡}
\author{Calin Miron‡}
\author{David Molnar*†}
\author{Brad Moon‡}
\author{Bruno Ostarcevic‡}
\author{Alberto Peltea‡}
\author{Matt Rosenberg‡}
\author{Catalin Sandu‡}
\author{Arthur Saputkin†}
\author{Sagar Shah*†}
\author{Daniel Stan‡}
\author{Ernest Szocs‡}
\author{Shengye Wan†}
\author{Spencer Whitman†}

\author{Sven Krasser‡}
\author{Joshua Saxe*†}

\abstract{
Today’s cyber defenders are overwhelmed by a deluge of security alerts, threat intelligence signals, and shifting business context, creating an urgent need for AI systems that can enhance operational security work. Despite the potential of Large Language Models (LLMs) to automate and scale Security Operations Center (SOC) operations, existing evaluations are incomplete in assessing the scenarios that matter most to real-world cyber defenders. This lack of informed evaluation has significant implications for both AI developers and those seeking to apply LLMs to SOC automation. Without a clear understanding of how LLMs perform in real-world security scenarios, AI system developers lack a north star to guide their development efforts, and users are left without a reliable way to select the most effective models. Furthermore, malicious actors have begun using AI to scale cyber attacks, emphasizing the need for open source benchmarks to drive adoption and community-driven improvement among defenders and AI model developers. 

To address this gap, we introduce CyberSOCEval, a new suite of open source benchmarks that are part of CyberSecEval 4. CyberSOCEval consists of benchmarks tailored to evaluate LLMs in two tasks: Malware Analysis and Threat Intelligence Reasoning, core defensive domains that have inadequate coverage in current security benchmarks.  Our evaluations reveal that larger, more modern LLMs tend to perform better, confirming the training scaling laws paradigm. We also find that reasoning models leveraging test time scaling do not achieve the boost they do in areas like coding and math, suggesting that these models have not been trained to reason about cybersecurity analysis, and pointing to a key opportunity for improvement.  Finally, we find that current LLMs are far from saturating our evaluations, demonstrating that CyberSOCEval presents a significant hill to climb for AI developers to improve AI cyber defense capabilities.

} 

\date{September 15 2025, Revised November 6 2025}
\correspondence{Lauren Deason at \email{laurendeason@meta.com}}
\metadata[Author Legend]
{* - Core Contributors
† - Meta
‡ - CrowdStrike}

\begin{document}
\maketitle

\section{Introduction}

\begin{figure}
\centering
\includegraphics[width=\textwidth]{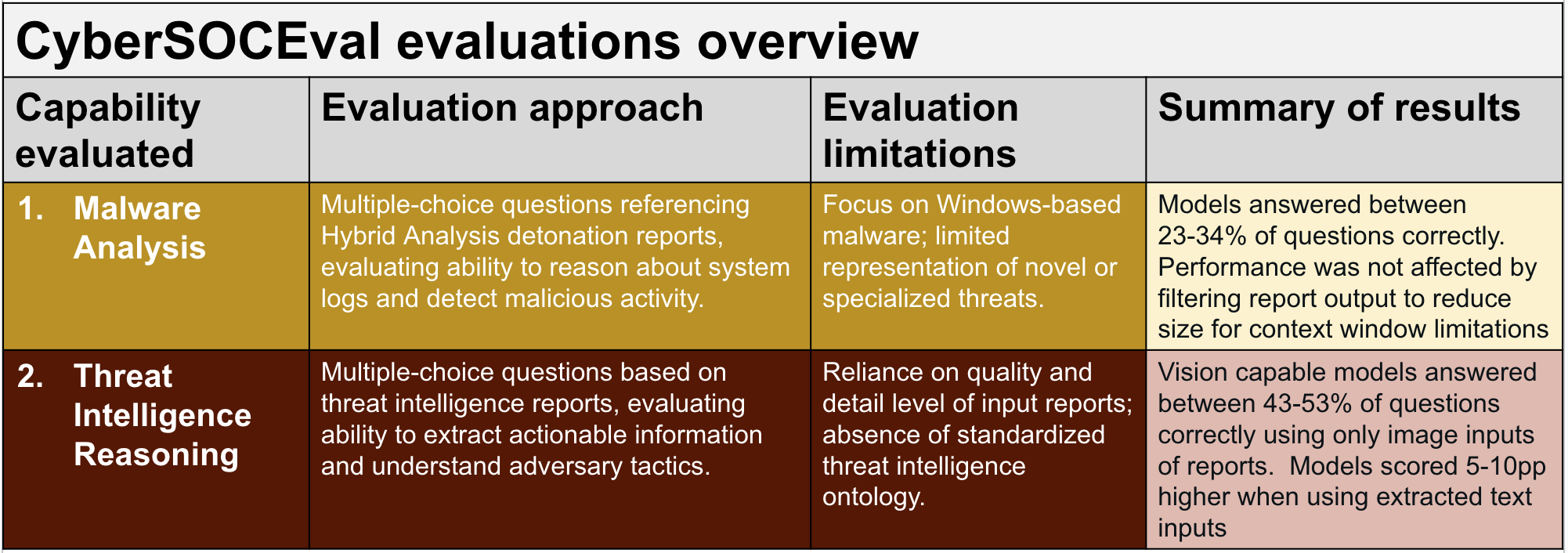}
\caption{Overview of CyberSOCEval, our proposed benchmark suite for evaluating Large Language Models (LLMs) in Security Operations Center (SOC) tasks. The suite consists of two complementary benchmarks: Malware Analysis and Threat Intelligence Reasoning, which together assess an LLM's ability to support key SOC functions.
}
\label{fig:overview}
\end{figure}

The growing complexity and volume of cyber threats have overwhelmed cyber defenders working in Security Operations Centers (SOCs). Large Language Models (LLMs) offer a promising avenue for automation, yet currently have skills gaps that hinder their utility, as we show below. Evaluations like those introduced here can guide model developers in improving their models’ cyber capabilities to better help cyber defenders.
Simultaneously, threat actors are using Artificial Intelligence (AI) to enhance their cyber attacks.  LLMs can generate highly convincing phishing emails, messages, and fake websites, expanding beyond attackers’ native languages.  AI systems also accelerate coding, serve as valuable technical assistants, and can aid in the research of evasion techniques. Malware developers benefit from these systems just as legitimate engineers do. Attackers have been observed to leverage all of these new AI capabilities in the wild, and we can expect continued attacker adoption of AI as capabilities grow (Machine Learning-Based Malware Detection in a Production Setting; sec. 5.5.2~\cite{Krasser2025malware}).

CyberSOCEval addresses the benchmark gap in cybersecurity, and contributes to supporting defenders as attack and defense becomes more AI-driven, by providing an open source benchmark suite that fosters wide access and community contributions.  Our suite of benchmarks leverages the experience of security practitioners at a hyperscaler technology company and a leading cyber breach prevention firm and evaluates LLM capabilities relevant to cyber defenders and SOC analysts, the first we are aware of that is open source. The main contributions of our work are:

\begin{itemize}
\item \textbf{Introduction of CyberSOCEval}: An open source cyber defense benchmark suite, unlike previous closed benchmarks, for AI systems. CyberSOCEval provides a North Star for AI model developers, and a selection metric for cyber practitioners looking to apply LLMs.

\item \textbf{Identification of a significant hill to climb for AI developers}: Our results show that current LLMs are far from saturating our evaluations, indicating a significant opportunity for improvement in cyber defensive capabilities while using our evaluations as a guide.

\item \textbf{Identification of performance themes across many popular LLMs}: We evaluate the performance of various LLMs on our benchmark suite, showing that larger, more modern models tend to perform better. Surprisingly, we did not observe a large performance boost from reasoning models, suggesting that targeted reasoning training in the cyber domain could be an area for model improvement.

\item \textbf{Value of our evaluations for practitioners}: The variance in results across models highlights the importance of our evaluations for helping practitioners decide which LLM to use.

\end{itemize}

The benchmarks we present focus on two key areas, chosen for their essential role in defensive operations according to subject matter experts on our team:
\begin{itemize}

\item \textbf{Malware Analysis}: Assesses the precision and recall of LLMs in identifying malicious activities from potential malware, such as detecting ransomware or remote access trojans. Accurate threat detection with minimal false positives is crucial for SOC analysts. Our evaluations index how well AI systems under test perform this currently heavily manual, expertise-dependent task. 

\item \textbf{Threat Intelligence Reasoning}: Evaluates an AI's ability to parse unstructured threat intelligence reports and extract actionable insights. SOC analysts rely on synthesizing threat intelligence to inform detection strategies, prioritize vulnerabilities, and understand adversary tactics, but threat intelligence feeds are often too voluminous for humans to process in detail. We test models’ ability to approximate this manual work, measuring their ability to map complex attack chains to frameworks like MITRE ATT\&CK and infer targeted industries.
\end{itemize}

The rest of this paper is structured as follows:

\begin{enumerate}
    
\item \textbf{Related Work}: We review existing security AI benchmarks and evaluations, highlighting the gaps that CyberSOCEval aims to fill. It covers malware analysis, and threat intelligence reasoning, both highly labor intensive tasks that require a high level of subject matter expertise.

\item \textbf{Benchmarks}: We provide an overview of the two benchmark components, detailing each benchmark’s format, evaluation criteria, and dataset construction.

\item \textbf{Experimental Setup and Results}: We give results for our evaluation across a range of popular LLMs, identifying key themes that explain variance in model performance.

\item \textbf{Discussion and Future Work}: We discuss the insights and observations from the results, including common failure modes, outlier models and the impact of filtering on telemetry data. We also discuss the implications for future research and development in AI-driven cybersecurity evaluations and capabilities.

\item \textbf{Conclusion}: The paper concludes with a summary of the findings.

\item \textbf{Appendix}: The appendix provides additional details on the methodology, including manual review protocol and the bootstrap approach for estimating benchmark power and minimum detectable effects.
\end{enumerate}

\section{Related Work}

\cite{sophos-benchmark}~included three fundamental SOC tasks to benchmark LLMs: (1) incident investigation (converting natural-language questions about telemetry into SIEM queries), (2) incident summarization from raw SOC data, and (3) incident severity rating.  The Sophos team did not open source their evals, but their technical report underscores the need for public SOC benchmarks, a gap that CyberSOCEval begins to help close.

The~\cite{oniagbi-evaluation} study of LLM agents uplift for Tier-1 SOC triage in a synthetic lab setting found that LLM assistance significantly accelerated initial triage and reduced analyst cognitive load, yet worked best as a co-pilot rather than an autonomous agent. This study also did not include an open source evaluation framework.

Because they usefully assess affordances and limitations of AI systems for SOC operators but do not release public benchmarks, the Sophos and Oniagbi studies underscore a need for targeted, open benchmarks, which CyberSOCEval provides.

 \cite{bhusal2024securebenchmarkinglargelanguage} targets cyber defense skills in the Industrial Control Systems domain, and does include an open benchmark; but because it focuses on a single industry vertical, it does not address the need for a more general benchmark suite.

CTIBench by ~\cite{alam2024ctibenchbenchmarkevaluatingllms} is a recent benchmark explicitly designed for cyber threat intelligence (CTI) tasks. It introduces a suite of four complementary evaluations covering threat-intel question answering and analytical reasoning: multiple-choice questions to probe threat knowledge, root cause mapping of vulnerabilities to underlying causes, vulnerability severity prediction using standards like CVSS, and threat actor attribution given attack descriptions.

\cite{ji2024sevenllmbenchmarkingelicitingenhancing} is a bilingual (English and Chinese) framework focused on threat intelligence and incident analysis. SEvenLLM curated a high-quality corpus of 8,485 threat reports to fine-tune domain-specific LLMs on 28 cybersecurity tasks, and compiled a 1,300-sample QA benchmark to evaluate CTI analytical capabilities.

CyberSOCEval goes beyond existing CTI LLM evaluation work by incorporating multimodal intelligence reports (e.g. combining textual indicators of compromise (IOCs) with tables or diagrams), which modern LLMs can now process, and provides higher-level reasoning test cases that current AI systems struggle with, but which are important manual bottlenecks for cyber defenders.

An evaluation of AI malware analysis from detonation reports has been missing. While often performed by specialized teams outside the SOC, malware analysis is important because it gives insights into an adversary’s potential activities on a compromised host. CyberSOCEval’s Malware Analysis suite is novel in benchmarking such abilities– it presents models with real sandbox outputs (e.g., process trees, network traffic from CrowdStrike Falcon® Sandbox) and poses questions that demand interpreting the malware’s behavior in context.

Beyond the specialized areas above, there have been broader efforts to benchmark LLMs on security knowledge and defensive reasoning.~\cite{wan2024cyberseceval3advancingevaluation} took a two-pronged approach: it assessed LLMs’ vulnerabilities (e.g. propensity for insecure code and malicious instruction-following) and their dual use capabilities like vulnerability identification.

Other works have constructed extensive question-answer banks to test factual security knowledge: for example, CyberMetric by~\cite{cybermetric} generated thousands of multiple-choice questions from standards (NIST, CWE, etc.) and evaluated dozens of models on this wide-ranging quiz.

In summary, existing security benchmarks either cover broad knowledge Q\&A or specific technical sub-tasks, and they highlight that general models often fall short on complex security reasoning. CyberSOCEval builds on this prior art by focusing more deeply on defender-centric reasoning in a SOC context. It complements general benchmarks with targeted tests that require \emph{context integration, multi-step reasoning, and judgement}.

\section{Benchmarks}

We selected two benchmark components - Malware Analysis Reasoning and Threat Intelligence Reasoning - to evaluate LLMs on a representative set of manual tasks that SOC operators and Blue Team operators engage in within their daily work.
We designed benchmarks to meet three criteria:

\begin{enumerate}
\item High performance should translate to real-world efficiency gains.
\item Evaluation should be fully automated.
\item The volume of questions should be sufficiently large to reliably detect moderate differences in model performance.
\end{enumerate}

Each of our benchmarks uses a multiple-choice question/answer format.
We ensured data quality through an iterative process of LLM generation, human review, and pipeline refinement, making manual edits to individual question-answer pairs and updating the pipeline to mitigate quality issues on each iteration, iterating until the outputs of the synthetic generation pipeline achieved our target threshold for quality.

\subsection{Malware Analysis Benchmark}

\begin{figure}
    \centering
    \includegraphics[width=\textwidth]{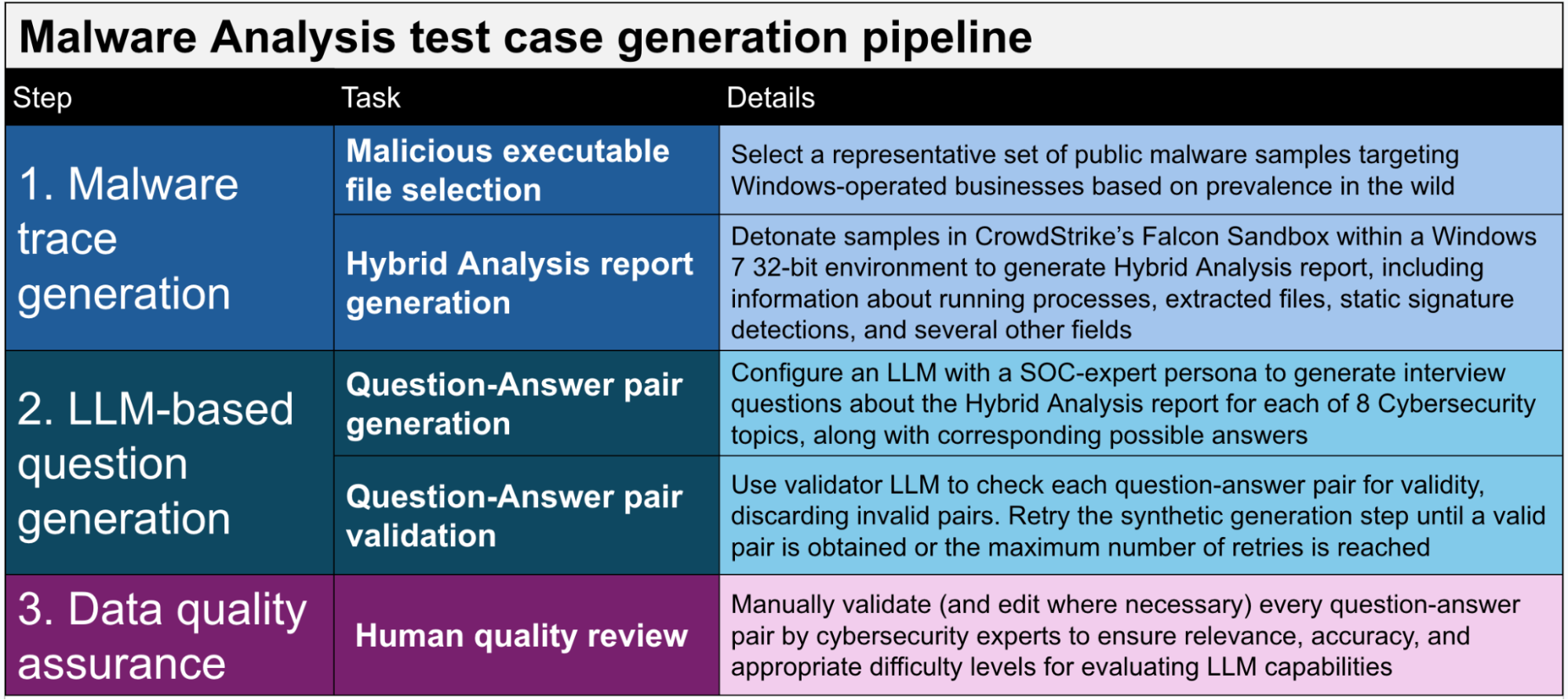}
    \caption{The steps followed for our Malware Analysis benchmark to generate and refine question-answer pairs using LLMs, and to manually validate.}
    \label{fig:b2_generation_overview}
\end{figure}

The Malware Analysis benchmark is designed to evaluate AI systems' ability to analyze complex, low level process execution data and understand potentially malicious signals. We derive our test cases from Hybrid Analysis report data produced by detonating samples in CrowdStrike Falcon® Sandbox. These logs include detailed information about running processes, extracted files, static signature detections, and more. The dataset covers five malware categories detailed in Table~\ref{tab:malware_family_label}.

\begin{table}
    \centering

    \begin{tabularx}{0.9\textwidth}{
  | >{\centering\arraybackslash}X 
  | >{\centering\arraybackslash}X | }
 \rowcolor{lightgray} Malware Type & Description \\
  \hline 
  \textbf{Endpoint Detection and Response/AntiVirus (EDR/AV) KILLERS} & Evades detection and disables security mechanisms on endpoints using techniques like process injection, obfuscation, persistence, and process termination. \\
  \hline
  \textbf{RANSOMWARE} &  Encrypts a victim's files, rendering the data inaccessible. Attacker then demands a ransom payment, typically in cryptocurrency, in exchange for the decryption key.  \\
  \hline
  \textbf{RAT - REMCOS Family} & Remote Access Trojan providing attackers with persistent remote access to a victim's computer using techniques including polymorphism, rootkit capabilities, vulnerability exploitation, and legitimate tool use. \\
  \hline 
  \textbf{INFOSTEALERS} & Infiltrates systems to extract sensitive information such as login credentials, financial information, etc. using techniques such as keylogging, for grabbing, and network traffic interception. \\
  \hline 
  \textbf{UM UNHOOKING} & Evades antivirus detection by manipulating the hooks that AV/EDR products use to monitor and control the execution of processes.  \\
  \hline 
  \end{tabularx}
    \caption{Five categories of malware from which samples were selected to be used as the basis for questions in our Malware Analysis benchmark.}
    \label{tab:malware_family_label}
\end{table}

\subsubsection{Test case structure and metrics}

Test cases task the system under test with analyzing JSON-formatted system log data from malware detonations and then answering multiple-choice questions, with up to ten possible correct answers, based on the detonation report.
Evaluation is based on accuracy: the share of questions for which the system selects all correct options and only the correct options.

\begin{figure}
    \centering
    \fbox{\includegraphics[width=0.9\textwidth]{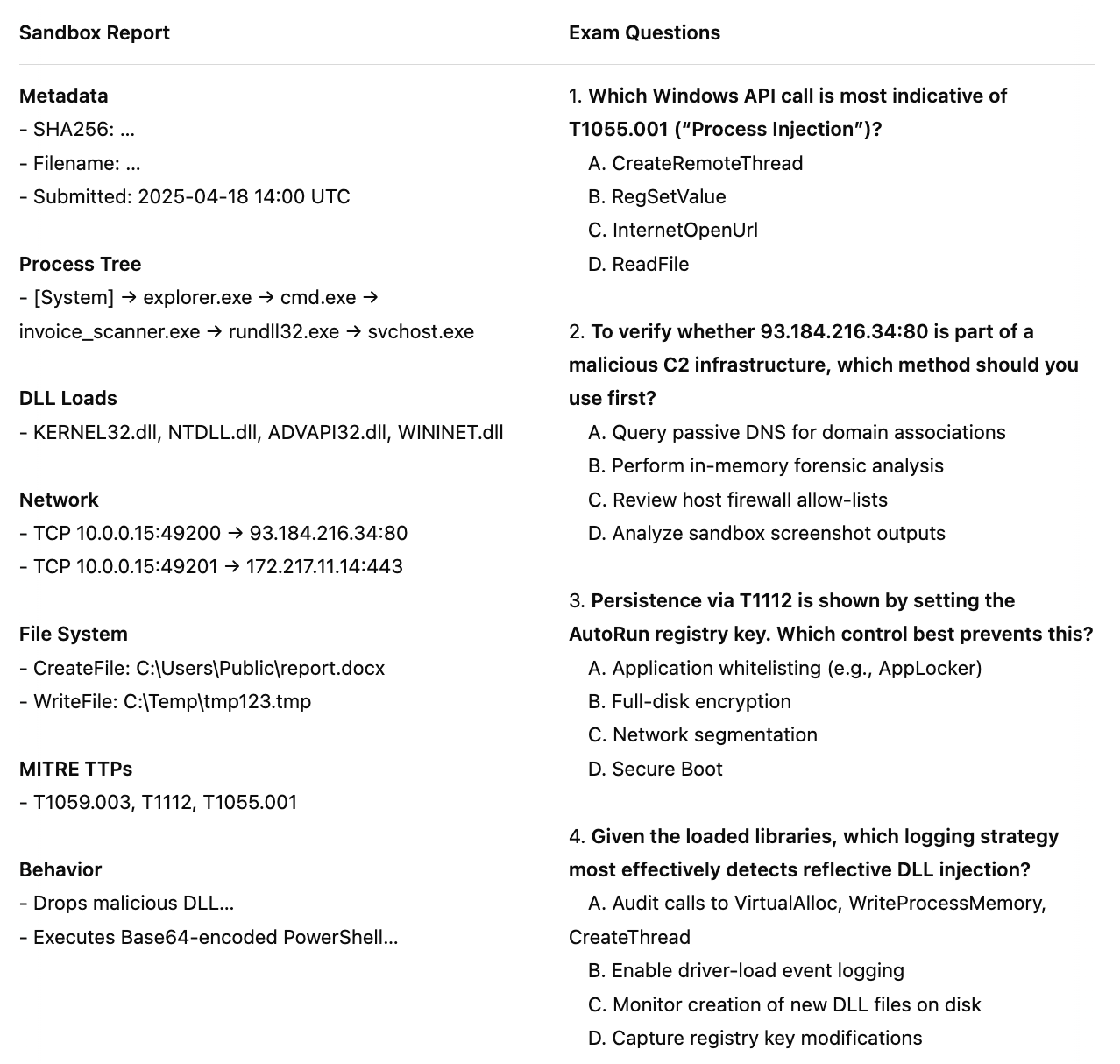}}
    \caption{ A notional, shorter example of our Malware Analysis test cases.  Our real test cases contain fully detailed sandbox detonation JSON files alongside multiple choice questions that exercise the AI system under test’s Malware Analysis abilities.}
    \label{fig:malware_analysis_examples}
\end{figure}

\subsubsection{Test case creation}

To create the 609 question and answer test cases in our evaluation, we selected public malware samples targeting Windows-operated businesses and detonated them in a controlled environment. Using Llama 3.2 90B, we synthetically generated interview questions and answers based on the Hybrid Analysis reports. Each question-answer pair was manually validated and edited by cybersecurity experts to ensure relevance, accuracy, and appropriate difficulty.

While the benchmark focuses on logs from malicious file detonations, we believe AI systems that perform well here can also reason about logs from non-file execution triggers. This potential for generalization is supported by the benchmark's coverage of attack categories and cybersecurity topics (see Section ~\ref{section:malware_analysis_coverage}). Future work may explore expanding the benchmark to include a wider variety of log sources and attack types, further enhancing its applicability and robustness.

\begin{figure}
    \centering
    \includegraphics[width=\textwidth]{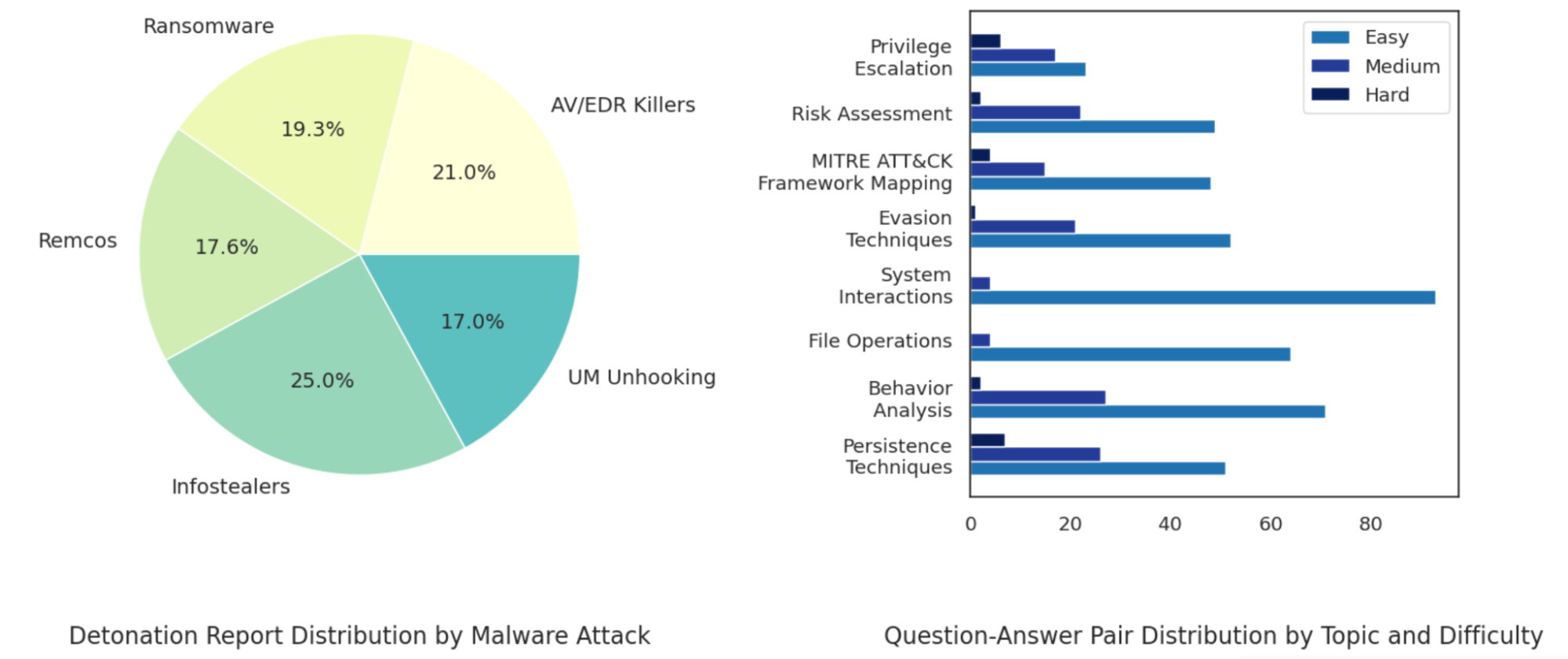}
    \caption{Distribution of malware families covered by detonation reports included in the Malware Analysis benchmark (left), and distribution of question-answer pairs by topic and difficulty rating (right).}
    \label{fig:b2_dataset_distribution}
\end{figure}

\subsection{Threat Intelligence Reasoning}

\begin{figure}
    \centering
    \includegraphics[width=\textwidth]{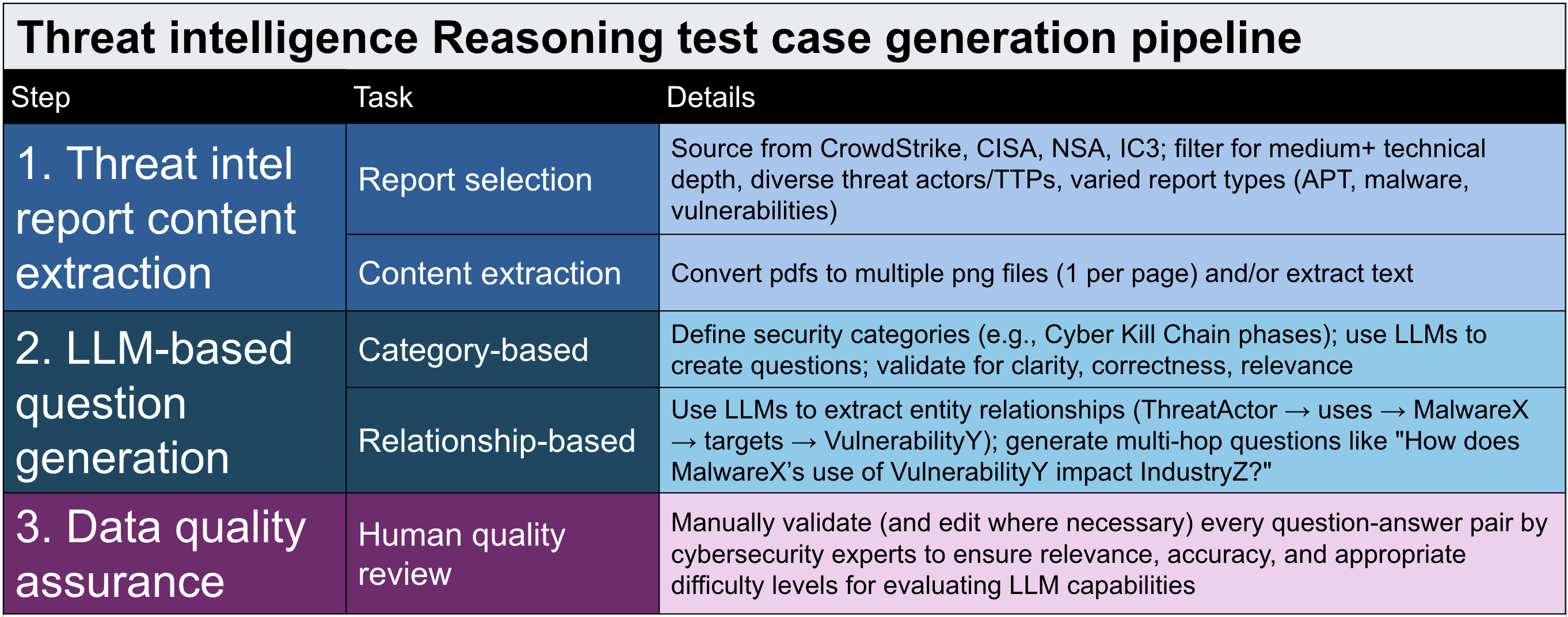}
    \caption{The steps followed for our Threat Intelligence benchmark to generate question-answer pairs using LLMs, and to manually validate.}
    \label{fig:b3_generation_overview}
\end{figure}

Our Threat Intelligence Reasoning benchmark is designed to evaluate an AI system's ability to understand, analyze, and extract actionable insights from threat intelligence reports. Unlike basic document comprehension tasks, this benchmark emphasizes security reasoning, such as identifying threat actors not explicitly mentioned, understanding complex attack chains, and mapping tactics to frameworks like MITRE ATT\&CK.

\subsubsection{Test case structure and metrics}

For each of our test cases, the AI system is provided with multiple images representing a threat intelligence report, with one image per report page. It must answer multiple-choice questions with up to six possible answers, where multiple options may be correct. Evaluation is based on accuracy: the share of questions for which the system selects all correct options and only the correct options.

\begin{figure}
    \centering
    \includegraphics[width=\textwidth]{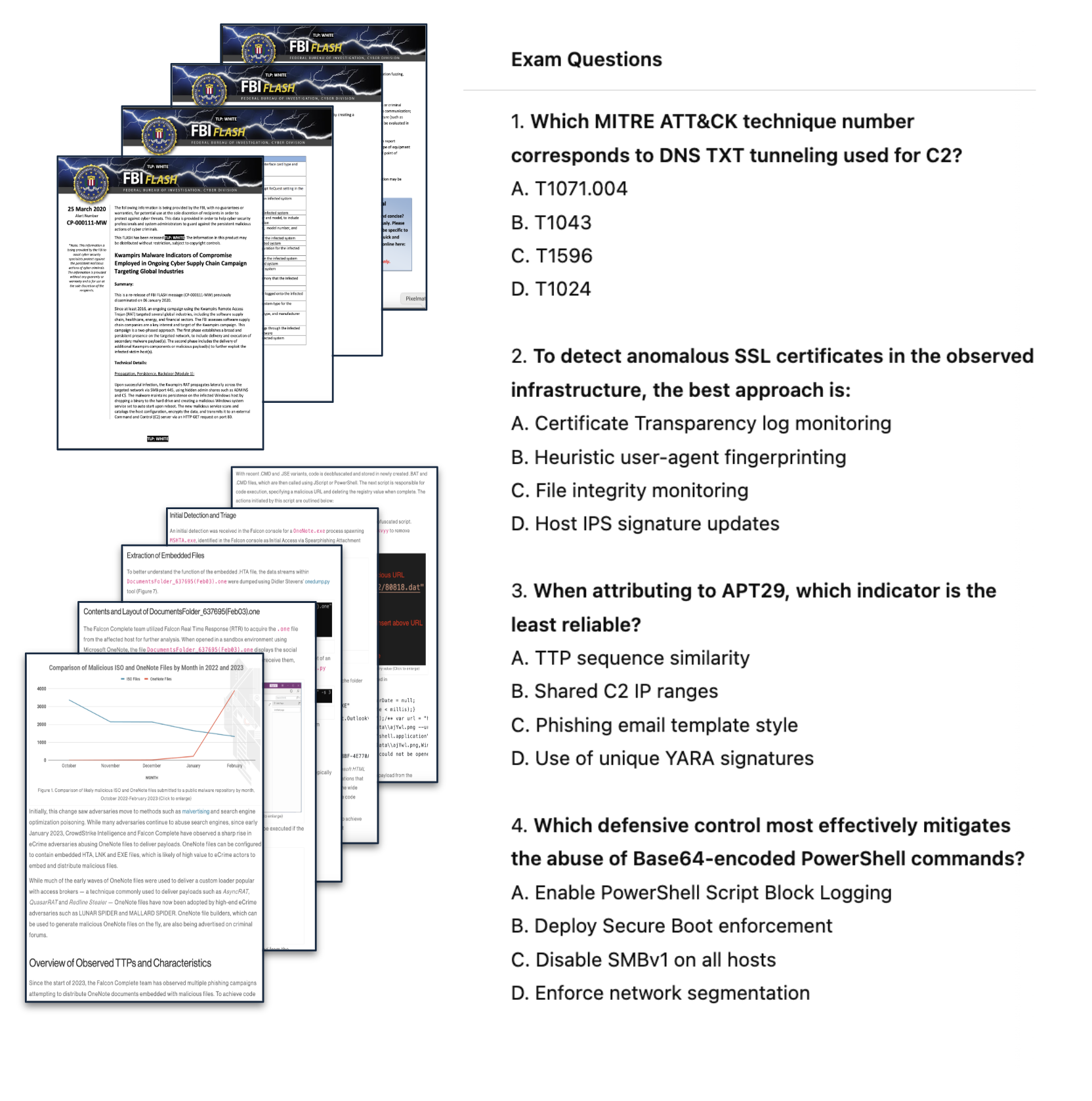}
    \caption{A notional, example test flow of our Threat Intelligence Reasoning tests.  We provide a threat intelligence report to the AI system under test via a set of images (one per report page), and then ask multiple choice questions that measure its ability to draw security-relevant conclusions.}
    \label{fig:threat_intelligence_example}
\end{figure}

\subsubsection{Test case creation}

We used Llama 4 Maverick and Llama 3.2 90B with two strategies for synthetically generating questions for these tests as follows:
\begin{enumerate}
\item \textbf{Category-Based Question Generation}: We began with a static list of manually defined categories, directing an LLM to produce questions about the extracted text from a given report, which were then validated for clarity, correctness, and relevance.
\item \textbf{Relationship-Based Question Generation}: We used an LLM to analyze the report text to generate a graph-like representation of key relationships (e.g., threat actors, malware, techniques), which we then sampled from and manually validated, iterating until question and answer quality met our thresholds for statistically significant test item accuracy.
\end{enumerate}

In addition to the synthetic data generation processes above, we also manually created a small number of questions whose answers depend entirely on charts or graphics in the report.

The final dataset consists of 588 question-answer pairs derived from 45 distinct threat intelligence reports. We included report types like APT campaign analyses, malware technical reports, and vulnerability advisories.

\begin{figure}
    \centering
    \includegraphics[width=\textwidth]{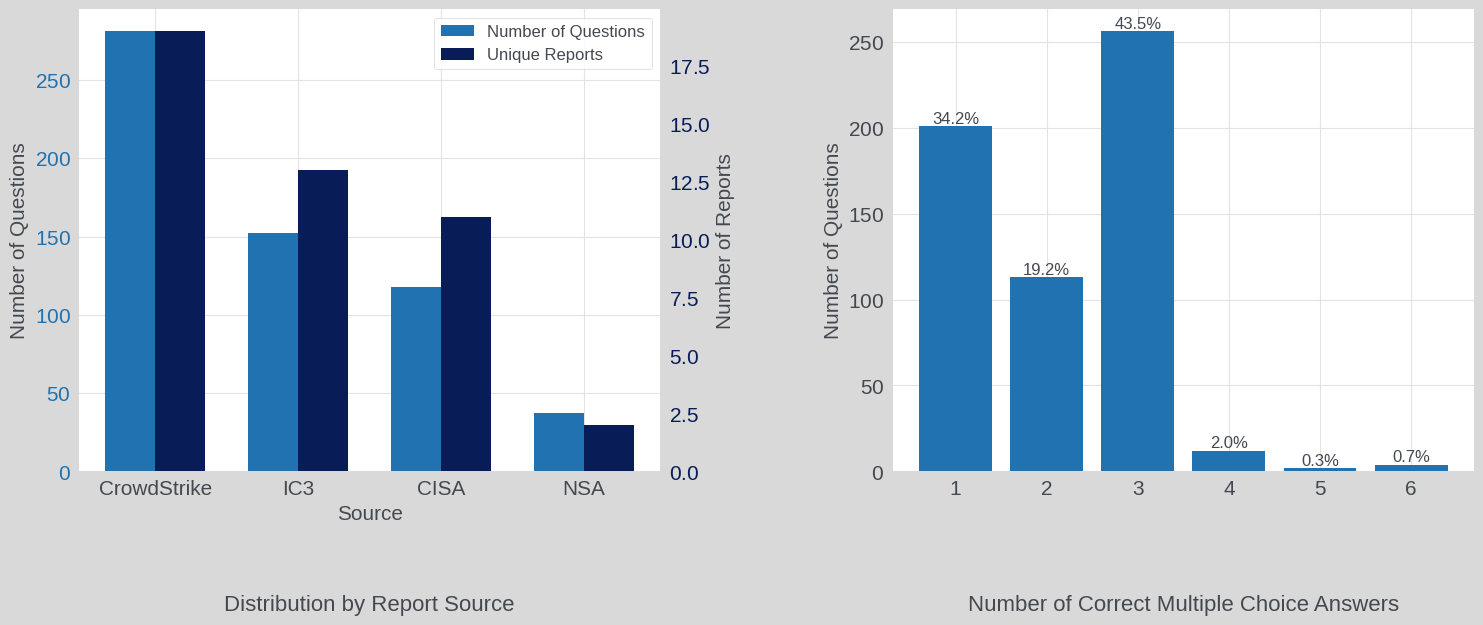}
    \caption{Number of questions (left-axis) or reports (right-axis) by report source (left).  Number of questions with the given number of correct multiple choice answers (right).}
    \label{fig:b3_dataset_distribution}
\end{figure}

\section{Experimental Setup and Results}

For each of the above benchmarks, we evaluate the performance of recent models from OpenAI, Google, Anthropic, Meta, and Deepseek. All models were tested using the original parameter values (in particular, no fine tuning was performed). Some manual adjustments to system prompts were made on a per-model basis to encourage the model-under-test to follow the desired response format in cases where a large share of responses failed to be parsed initially. All hyperparameter settings (e.g. temperature, top\_p, and 'thinking level' settings) remained at the default values set by the provider.

\subsection{Malware Analysis}
\label{sec:b2_results}

\begin{figure}
    \centering
    \includegraphics[width=\textwidth]{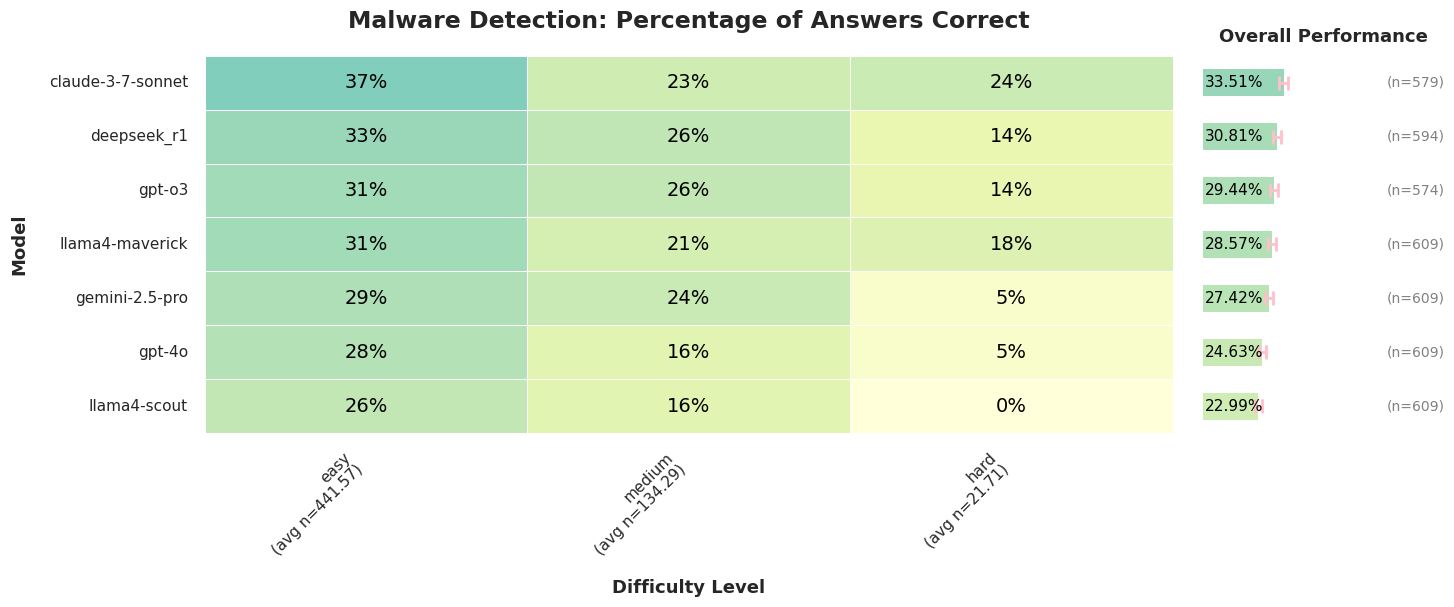}
    \caption{Share of completely correct multiple choice responses (Accuracy) for each model under test against the Malware Analysis benchmark, broken out by difficulty level. Average number of questions with parsable responses within each difficulty level reported as avg n for each column, total parsable responses in each row.  95\% confidence intervals shown in pink. }
    \label{fig:b2_results_by_difficulty_level}
\end{figure}

For each model evaluated against the Malware Analysis benchmark, Figure~\ref{fig:b2_results_by_difficulty_level} presents the share of questions that the model answered correctly broken out by question difficulty rating. The overall accuracy scores range from approximately 23-34\%, which should be compared to a baseline of approximately 0.63\% accuracy that would be expected from purely random guessing (see Appendix~\ref{sec:appendix_b2_baseline}).

Given the large size of the Hybrid Analysis reports, which was often near or in excess of the 128,000 tokens context window limit of several models under test, we also explored whether the entirety of the Hybrid Analysis report was needed to achieve the given performance level.  We found that in general, an abridged version of the report does not meaningfully affect performance, and so models with smaller context windows may be evaluated on this benchmark as well, by setting a flag to filter the input report.  More detail on the logic used to reduce the report size can be found in Appendix~\ref{sec:appendix_b2_context_size}, along with evidence that this filtering has very little effect on benchmark performance.

Additional summary statistics of model performance on this benchmark including breakouts by malware type and question topic, as well as Jaccard similarity score (which allows partial credit for multiple choice selections that overlap imperfectly with the correct set of answers) can be found in Appendix~\ref{sec:appendix_b2_additional_Results}.

\subsection{Threat Intelligence Reasoning}
\label{sec:b3_results}

\begin{figure}
    \centering
    \includegraphics[width=\textwidth]{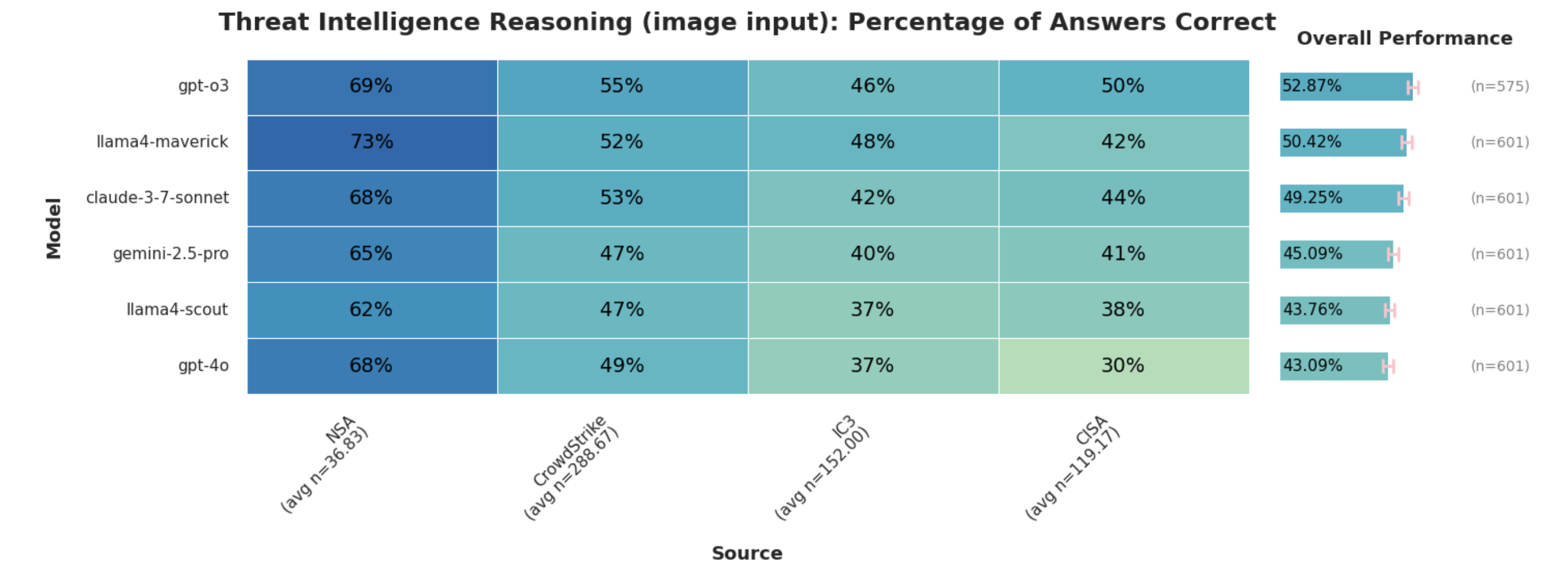}
    \caption{Per model evaluation scores for percent of multiple choice questions answered with perfect accuracy in the Threat Intelligence Reasoning benchmark, broken out by source of the Threat Intel report on which the questions are based. Average number of questions with parsable responses within each source category reported as avg n, total parsable responses in each row.  95\% confidence intervals shown in pink. }
    \label{fig:b3_results_by_source}
\end{figure}

For each model evaluated against the Threat Intelligence Reasoning benchmark, Figure~\ref{fig:b3_results_by_source} presents the share of questions the model correctly answered broken out by the source of the Threat Intelligence report on which the questions were based.  Overall scores range from approximately 43\%-53\%, which should be compared to a baseline of approximately 1.7\% accuracy that would be expected from purely random guessing (see Appendix~\ref{sec:appendix_b3_baseline}).

Additional summary statistics of model performance on this benchmark including Jaccard similarity score (which allows partial credit for multiple choice selections that overlap imperfectly with the correct set of answers) can be found in Appendix \ref{sec:appendix_b3_additional_results}.  Additionally, a comparison of model performance on this benchmark when the model is provided the Threat Intel reports as images versus extracted text is presented in the discussion section below.

\section{Discussion}

Given the results from the previous section, we discuss here in more detail some of the interesting patterns observed in model responses and implications for further development and use of AI systems in security operations.

\subsection{Malware Analysis Coverage}
\label{section:malware_analysis_coverage}

We argue that performance on our malware analysis data benchmark provides a reasonable proxy for an AI system’s capabilities to reason about host behaviors that are triggered not only by malicious executable files, but that this also generalizes to detection and reasoning capabilities about logs generated from fileless attacks.

As background, system logs are automatically generated records containing details about events that occurred in a given host environment. These include logs of system actions and modifications, and can be categorized into two primary types based on their generation trigger: file-based logs and fileless logs. File-based generated logs represent events that were triggered by file execution, while fileless logs represent events that were triggered by some other action, such as direct execution of commands inside a terminal, or manually downloading an attachment from an email.

While there exist some system events that can only by triggered by fileless actions, and others that can only be triggered by file-based, the majority of system events can appear as a result of either a file-based or fileless trigger, and will appear with the same system footprint in logs, regardless of initial trigger type. For example:

\begin{itemize}

\item Direct execution of commands inside a terminal can be done programmatically via executables
\item Manual Double-click of an executable can be replaced with an execution command
\item Manual downloading can be replaced with a GET request
\end{itemize}

Because of this overlap, we believe that our Malware Analysis benchmark provides signal on detection capabilities over both malware and fileless attacks, despite the fact that the system logs in our dataset are triggered by file execution. 

While we believe that our dataset provides a good proxy for capabilities to reason about these types of attacks, we recognize that there are additional types of attacks that may be high priority for an organization which may not be covered by the system event logs used in our benchmark, such as IP exfiltration or accessing of sensitive assets. We hope to expand our dataset in the future to include additional logging to cover a broader scope of attacks, as well as to explicitly include system logs from fileless attack simulations.

\subsection{Multimodal Threat Intelligence Reasoning}

The inclusion of multimodal threat intelligence reports (e.g. combining textual IOCs with tables or diagrams) in this benchmark aims to address a gap over existing frameworks like CTIBench by ~\cite{alam2024ctibenchbenchmarkevaluatingllms} and SEvenLLM by~\cite{ji2024sevenllmbenchmarkingelicitingenhancing}. Our approach better reflects the reality of how threat intelligence is communicated in practice, where critical information is often distributed across text, tables, diagrams, and other visual elements.

\begin{figure}
    \centering
    \includegraphics[width=\textwidth]{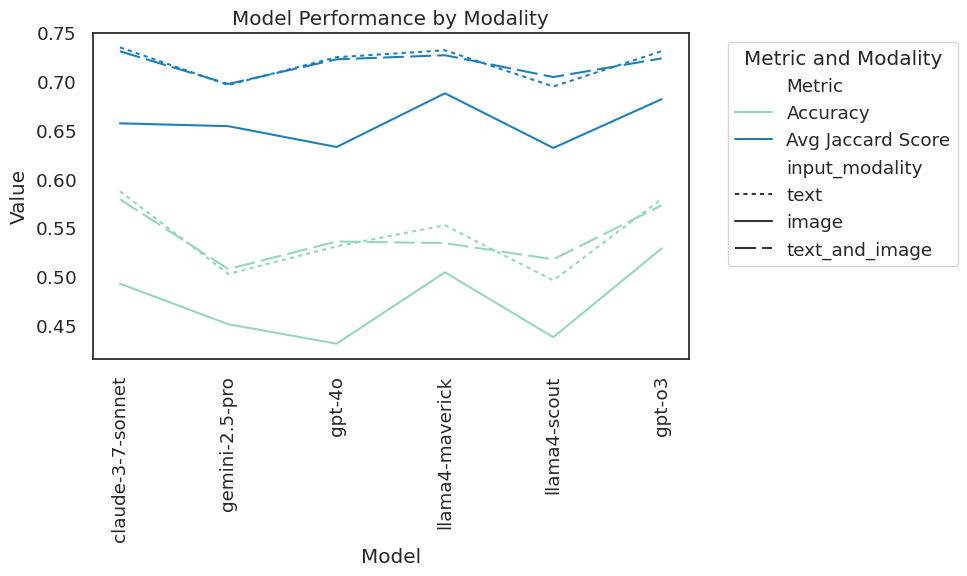}
    \caption{Accuracy (\% answers correct) and average Jaccard Score (overlap between model answers and correct answer set) on the Threat Intelligence Reasoning Benchmark for models when provided with the content of the report as text or as images}
    \label{fig:b3_claude_by_modality}
\end{figure}

The results for this benchmark presented in Section~\ref{sec:b3_results} show how various models score when provided the content of the Threat Intel report as multiple images. Here, we additionally look at the relative performance of these models on the benchmark when provided with the report content as extracted text instead of images, as well as with both text and image. In Figure~\ref{fig:b3_claude_by_modality}, we see that for all models in our experiment, extracted text yields better performance than image inputs alone. Providing the model with both extracted text and image inputs appears to yield similar, and in some cases even slightly worse, performance than text alone.

\begin{figure}
    \centering
    \includegraphics[width=\textwidth]{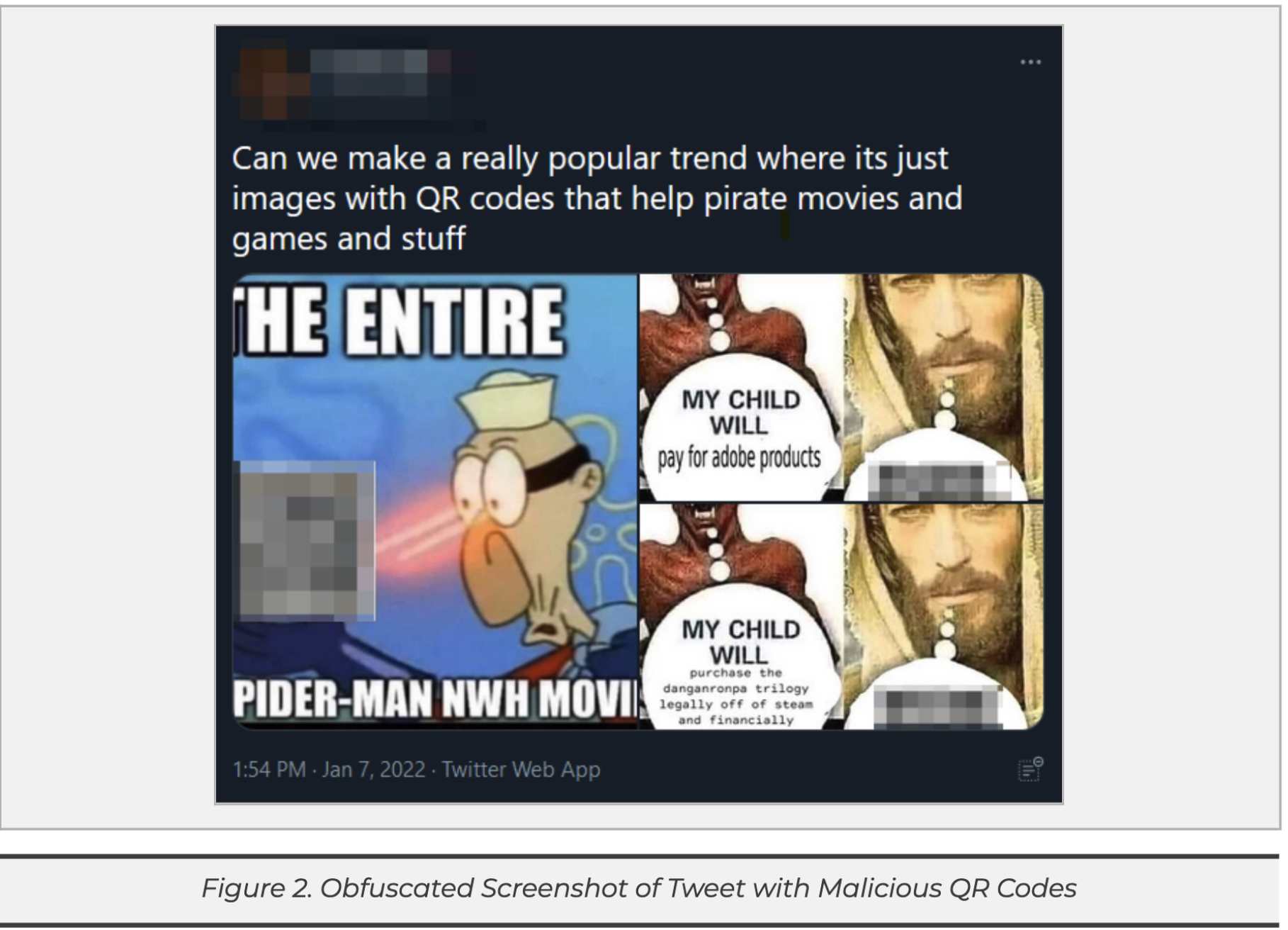}
    \caption{Image included in threat report at~\cite{csit-2311} which is needed to correctly answer a benchmark question about the initial access vector used by the threat actor. All six of the models under test answer correctly when provided with the report in image form, while only one model (o3) is able to answer correctly when receiving only text input.  Interestingly, when models are given both text and image input simulataneously, only half of them (o3, 4o, and gemini-2.5-pro) answer correctly, indicating that the models may be overly reliant on text inputs when provided multiple modalities.}
    \label{fig:image_example}
\end{figure}

The fact that vision capable models appear to do better on this benchmark using text input vs image input indicates that there is progress to be made for multi-modal models at reasoning about text ingested via other modalities. It also indicates that our dataset is biased towards text-based information, though it does also contain a small number of manually crafted questions to test analysis capabilities on purely visual inputs (see Figure~\ref{fig:image_example}).

The ability to process and reason across different information modalities may be as important as pure reasoning capabilities, particularly for organizations that rely on diverse threat intelligence sources posing direct implications for SOC teams selecting models for threat intelligence tasks. Future model development should prioritize robust multimodal understanding capabilities alongside improvements in security-specific reasoning to address the real-world bottlenecks faced by cyber defenders who currently must manually integrate information across these different formats.

\subsection{Impact of Inference-Time Thinking on Performance}
\label{subsec:inference_time_thinking}

Comparing the performance of reasoning models that leverage inference-time thinking (e.g. o3, deepseek-r1) against other models, we were initially surprised to find that we did not observe a more dramatic impact on performance of general reasoning capabilities akin to the ‘step-change’ in performance of reasoning models that has been observed in other domains such as coding and mathematics. We hypothesize that while test-time thinking may play an important role for defensive security outcomes, there may also be a need for more targeted training using security relevant data and reasoning traces. We believe this presents an opportunity for researchers and model developers to experiment with training techniques that might improve the fundamental reasoning capabilities of models in the security domain. 

\subsection{Additional Limitations}

We acknowledge that the use of multiple choice format in our evaluations does not provide a perfect proxy for capabilities to perform real world tasks that might require an AI system to provide free-form responses and to decide when actions need to be taken. We elected to use this format in order to maximize repeatability and interpretability of benchmark results, at the cost of some reduction in realism. We attempt to mitigate this by allowing for a relatively large number of multiple choice options with multiple correct answers (reducing the baseline probability of correct guessing).

It is important to also recognize the existence of performance bias that may be introduced in cases where the model under test is the same, or has similarities with the set of models that were used in synthetic data generation pipelines.  To ensure compliance with third party API acceptable use policies, we restricted our synthetic data generation pipelines to use a combination of Llama 3 and Llama 4 models, which could artificially boost the apparent performance of these models on the benchmarks.  We attempted to mitigate this concern through a combination of complex multi-agent generation pipelines combined with extensive manual review and editing, however it remains the case that this is a potential source of bias.

Another important limitation to call out is that the contextual data provided to an AI system under test in each of our benchmarks may differ significantly from the contextual data that a given business or security operations center observes in production. For this reason, it is important to keep in mind that performance of an AI system on one of our benchmarks (as with any benchmark) should not be interpreted literally as a prediction of the performance one would expect to observe in production for a given metric (e.g. accuracy). 

While our datasets are sourced to cover a representative set of attacks commonly seen in the wild and commonly used threat intelligence sources, for any particular organization, the distribution of likely attacks, or relevant threat intelligence, may differ significantly from the distribution in this test set.

While the performance of a given system on our benchmarks cannot be used to predict performance metrics in a specific organizational setting, it can provide a relative signal about the fundamental reasoning capabilities of any two systems at making effective security decisions.  Our intent is that these benchmarks provide a ‘hill to climb’ for model developers such that future generations of models improve upon current models at these fundamental security reasoning skills. Individual organizations will likely still wish to develop internal test sets to measure the performance of AI systems on their specific production data distributions, and may also benefit from further fine-tuning base models towards specific desired security capabilities.

\subsection{Future Extensions}

Looking ahead, we see several promising directions for enhancing and expanding CyberSOCEval benchmarks:

\begin{enumerate}
\item \textbf{Additional attack coverage}: While we believe that our existing Malware Analysis benchmark provides a reasonable proxy for an AI system’s capabilities to reason about host behaviors that are triggered by both malicious executable files and fileless attacks, we aim to expand the benchmark in the future to include explicit examples of fileless attack telemetry. In addition, we hope to expand coverage to other types of malicious activity, such as sensitive data exfiltration.

\item \textbf{Expansion to additional environments}: Our current Malware Analysis benchmark is specific to a model’s capability to reason about system logs in a Windows environment.  In the future, we aim to expand this dataset to include example logs and corresponding questions from additional environments such as Linux and mobile.

\item \textbf{Expansion of image modality dependencies}: While our Threat Intelligence Reasoning benchmark tests the capability of the system under test to reason about information provided through text and image modalities, our questions do not yet place a heavy emphasis on the ability to integrate information from charts, figures, and other exclusively non-text modalities. This is due to the fact that the majority of the content in our set of threat intelligence reports, as well as our synthetic question-answer generation logic is text based. In the future, we aim to increase the coverage in our benchmark of more diverse content within reports, as well as the inclusion of questions that require the system under test to rationalize jointly about text and non-text inputs in order to succeed.

\item \textbf{Standardized Threat Intelligence Ontology}: In the future, we aim to extend our threat intelligence reasoning benchmark beyond the multiple choice questions to include entity extraction and relationship mapping tasks. Models that are able to perform well on these tasks could then be trusted to automate structured data ingestion from threat intelligence reports, including entities that are not explicitly mentioned in the reports. Two obstacles that we faced in our initial attempt to include such a task were a) aligning on the ontology to be used for this task, and b) subjectivity of ground truth for mapping information in a given report to this ontology. These challenges highlight a fundamental issue in the cybersecurity field: the lack of universally accepted frameworks for describing and relating security concepts beyond MITRE ATT\&CK, which hinders real-world threat intelligence sharing and automation. In the future, we hope to engage with the community to develop more consistent ontologies or adapt to multiple competing frameworks, improving standardization across the industry.

\item \textbf{Additional benchmark to measure Incident Response and Triage Capabilities}: In future work, we hope to complement the benchmarks above with an evaluation that assesses a model's ability to correctly respond to incidents, prioritize threats, and identify appropriate mitigation within a specific business context.  A key difficulty here will be the subjective nature of how SOC analysts should 'correctly' prioritize different threats within their organization, and providing adequate examples of 'business context' to a model-under-test that are both comprehensive and realistic.

\item \textbf{Assessing recent Frontier and Security Focused Models}: Since the research on this project was completed, several new models have been released (eg Claude 4, GPT-5, etc) as well as some models that are specifically trained to maximize security knowledge and capabilities (eg Foundation-Sec-8B).  We are interested to see how these models perform on our benchmarks, and more generally look forward to seeing what findings come out of the research community using these benchmarks.
\end{enumerate}

\section{Conclusion}

We introduced CyberSOCEval, to our knowledge the first open source benchmark for LLMs focused on representative SOC activities. 

Our CyberSOCEval benchmark suite will be beneficial in addressing the critical evaluation gap in modern cybersecurity operations. As both cyber attacks and defenses increasingly leverage artificial intelligence, security teams require robust tools to assess and enhance their capabilities. This benchmark suite fills that need by providing freely accessible benchmarking tools while encouraging community participation and contribution, ensuring the framework remains current and relevant in the rapidly evolving threat landscape. Through its collaborative approach, CyberSOCEval enables security professionals to effectively measure, compare and improve their defensive capabilities against emerging AI-driven threats.

We showed that the benchmark is not saturated by any current LLM, indicating that more research is needed and desired to tune LLMs for cybersecurity defense. We encourage others to build on our results by adding additional tasks to CyberSOCEval.

\section{Acknowledgements}

The authors would like to thank Faizan Ahmad, Will Alexander, Justin Anderson, Adam Bradbury, Kevin Castillo, Stefan Cicos, Todd Fletcher, Chad Greene, Josh Grunzweig, Marian Gusatu, Adam Hannibal, Atharav Hedage, Randy Heins, Pete Huitsing, Amir Jalali, Pranau Kumar, Sherry Yikun Liu, Jake Lomas, Maroof Mansuri, Clay Moody, Asit More, Dan Pistelli, Alex Preneta, Chris Rohlf, Josh Ryder, Suraj Sawant, Valeriu Statache, Stefan Stein, Razvan Stoleriu and Dimitrie-Octavian Valu for their input and expertise. The authors would further like to thank Mihai Alestar, Liza Bales, Noah Bason, Roxana Boriceanu, Shelby Mann Bryant, Nicole Catalano, Brian Clerkin, Alexandra Damir, Bysshe Easton, Faith Eischen, Jared Engstrom, Camille Stewart Gloster, Vincent Gonguet, Jim Gust, Kat He, Mike Hollander, Cheryl Houser, Bilal Issa, Klaudia Krawiecka, Nicholas Longenbaugh, Will Lowe, Bhavesh Mehta, Adam Meyers, Jinpeng Miao, Ayaz Minhas, Abraham Montilla, Ionut Negru, Anukrati Omar, Melody Petersen, Cristian Popa, Tatyana Poturnak, Andrei Preda, and Rahul Verma for their help and support.

Additionally, the authors are extremely grateful to Anthony Penta, Ram Shankar Siva Kumar, and the Microsoft Security AI cyber evaluation group for reporting a bug in the Malware Analysis benchmark code.  Based on their findings, we have corrected the bug, and updated the results in the paper for this benchmark in October 2025.

\newpage
\bibliographystyle{plainnat}
\bibliography{cybersoc}

\newpage

\setcounter{section}{0}
\renewcommand{\thesection}{\Alph{section}}
\input{appendix}

\end{document}

%% file: appendix.tex
\section{Malware Additional Results}
\label{sec:appendix_b2_additional_Results}

\begin{figure}
    \centering
    \includegraphics[width=\textwidth]{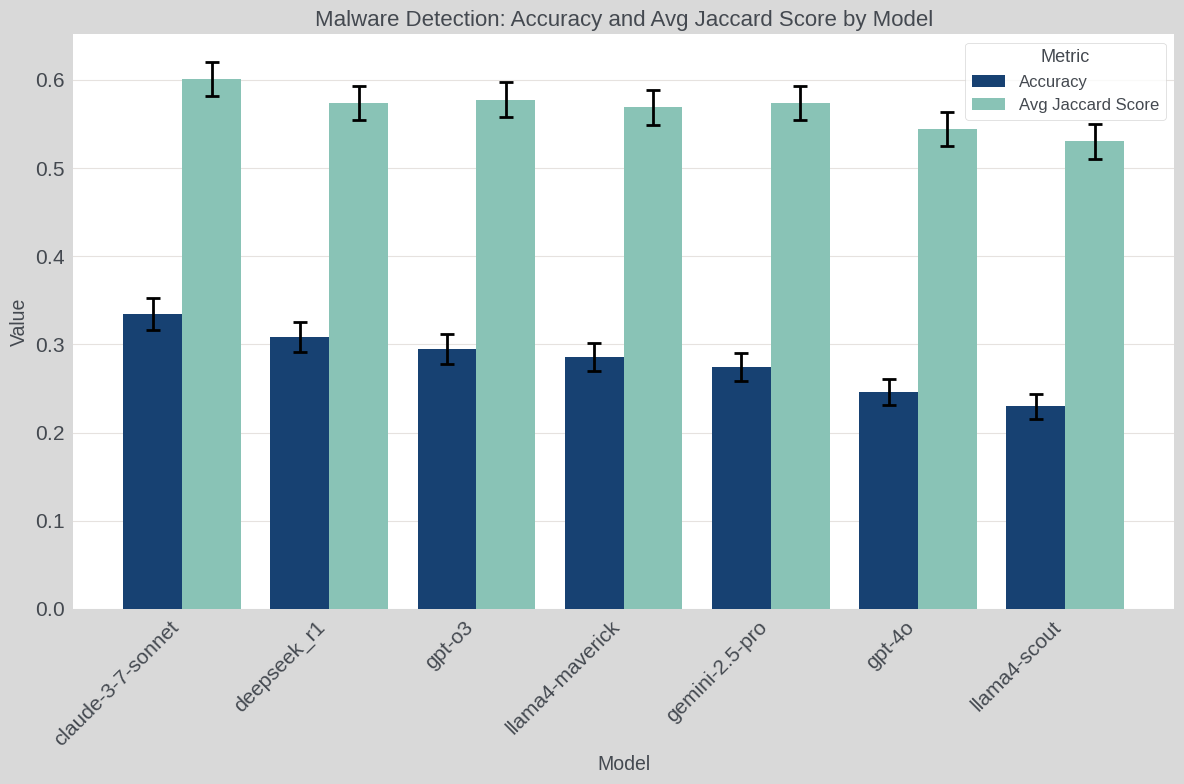}
    \caption{Share of completely correct multiple choice responses (Accuracy) and average Jaccard Score (over set of model provided responses and correct responses to multiple choice questions) for each model under test against the Malware Analysis benchmark.  Error bars reflect 95\% confidence intervals computed using a normal approximation.}
    \label{fig:b2-topline-results}
\end{figure}

\begin{figure}
    \centering
    \includegraphics[width=\textwidth]{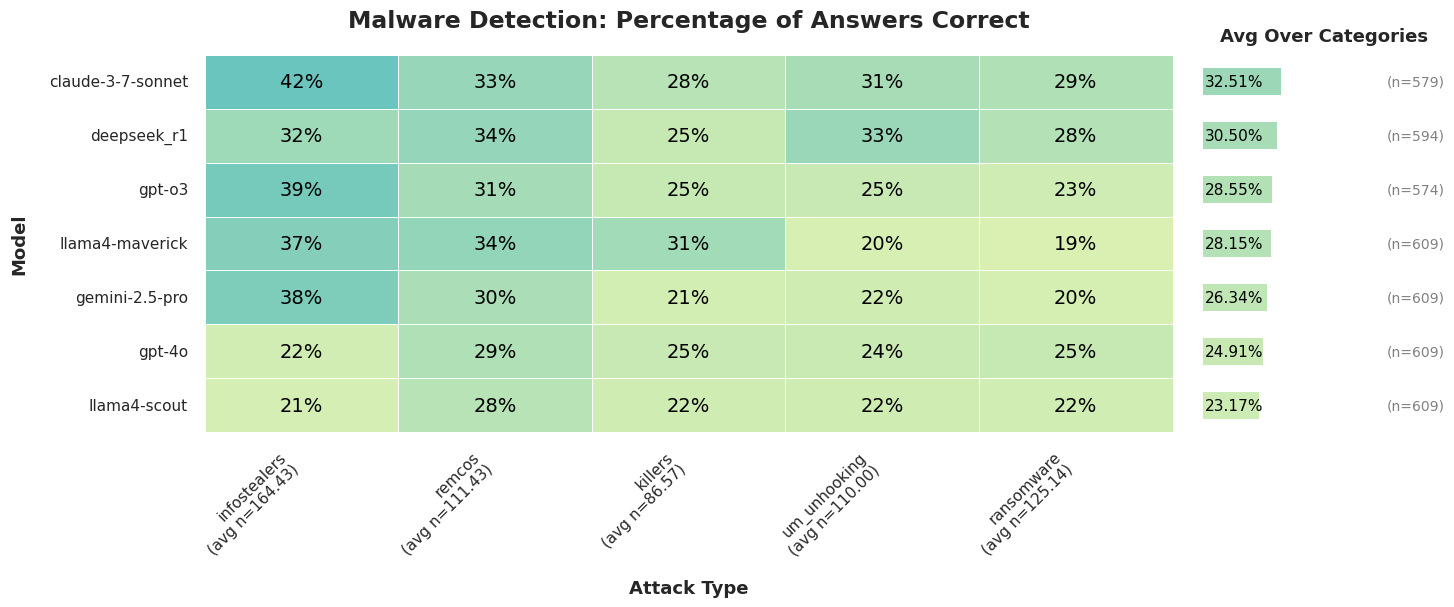}
    \caption{ Share of completely correct multiple choice responses (Accuracy) for each model under test against the Malware Analysis benchmark, broken out by malware family.   Average number of questions with parsable responses within each topic reported as avg n, total parsable responses by model in each row.}
    \label{fig:b2-results-by-malware-family}
\end{figure}

\begin{figure}
    \centering
    \includegraphics[width=\textwidth]{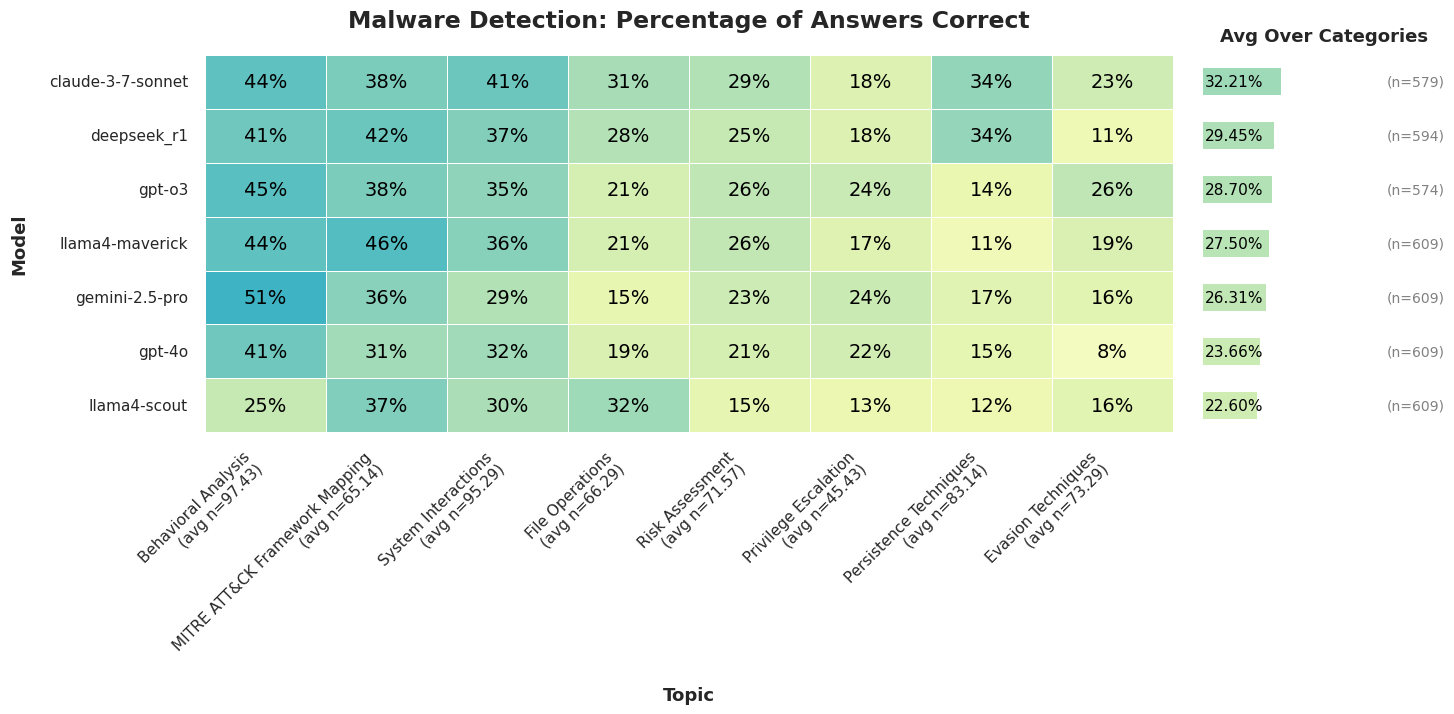}
    \caption{Share of completely correct multiple choice responses (Accuracy) for each model under test against the Malware Analysis benchmark, broken out by question topic.   Average number of questions with parsable responses within each topic reported as avg n, total parsable responses by model in each row.}
    \label{fig:b2-results-by-topic}
\end{figure}

Figure~\ref{fig:b2-topline-results} shows the share of completely correct multiple choice responses (Accuracy) and average Jaccard Score (over set of model provided responses and correct responses to multiple choice questions) for each model under test against the Malware Analysis benchmark. Figure~\ref{fig:b2-results-by-malware-family} shows the share of completely correct multiple choice responses (Accuracy) for each model under test against the Malware Analysis benchmark, broken out by malware family. Figure~\ref{fig:b2-results-by-topic} shows the share of completely correct multiple choice responses (Accuracy) for each model under test against the Malware Analysis benchmark, broken out by question topic.

\FloatBarrier

\section{Baseline Accuracy Computation for Malware Detection Benchmark}
\label{sec:appendix_b2_baseline}

The baseline accuracy of a random guesser on the Malware Analysis benchmark depends on the exact strategy taken by this random guesser, but would be expected to range between .63\% and 4.3\% depending on whether/how the guesser randomly chooses the number of options they will randomly select.

The probability of correctly guessing the exact set of correct answers for a question with K correct answers (assuming that the guesser's strategy is to first select uniformly the total of number of options they will select) is as follows:
\[
\Pr(\text{perfect}\mid K)
   \;=\; \frac{1}{9}\,\frac{1}{\binom{9}{K}},
   \qquad K=1,\dots,9.
\]

The expected accuracy over the whole test under this guessing strategy is:
\begin{align*}
\text{Expected accuracy}
  &= \sum_{K=1}^{9} p_K\;
     \Pr(\text{perfect}\mid K)                                  \\[4pt]
  &= \sum_{K=1}^{9}
     \frac{p_K}{9\,\binom{9}{K}}                               \\[6pt]
  &= \frac{0.3882}{9\cdot 9}
   + \frac{0.2319}{9\cdot 36}
   + \frac{0.1727}{9\cdot 84}
   + \frac{0.1168}{9\cdot 126}                                 \\[2pt]
  &\quad
   + \frac{0.0411}{9\cdot 126}
   + \frac{0.0263}{9\cdot  84}
   + \frac{0.0115}{9\cdot  36}
   + \frac{0.0099}{9\cdot   9}
   + \frac{0.0016}{9\cdot   1}                                 \\[6pt]
  &\approx 0.00625.                                            
\end{align*}

\[
\boxed{\text{Expected accuracy}\;\approx\;0.63\%}
\]

If instead the guesser chooses only a single option, chosen unifomly at random from the 9, only questions with exactly one correct answer (K=1) can be answered perfectly and the overall expected accuracy of this strategy would be:

\[
\Pr(\text{perfect match})
   \;=\; p_1 \times \frac{1}{9}
   \;=\; \frac{236}{608}\times\frac{1}{9}
   \;\approx\; 0.04313.
\]

\[
\boxed{\text{Expected accuracy}\;\approx\;4.3\%}
\]

\section{Baseline Accuracy Computation for Threat Intelligence Benchmark}
\label{sec:appendix_b3_baseline}

The baseline accuracy of a random guesser on the Threat Intelligence benchmark depends on the exact strategy taken by this random guesser, but would be expected to range between 1.74\% and 2.9\% depending on whether/how the guesser randomly chooses the number of options they will randomly select from.

Assuming that the naive guesser chooses uniformly the size of the answer set, given the distribution of correct number of answers (see ~\ref{fig:b3_dataset_distribution}), the baseline performance on this benchmark is as follows:

\[
\Pr(\text{exact match}\mid K)
=\frac{1}{6}\cdot \frac{1}{\binom{6}{K}}.
\]
\[
\mathbb{E}[\text{accuracy}]
=\sum_{K=1}^{6} p_K\cdot \frac{1}{6\binom{6}{K}}
=\frac{201}{588\cdot36}
+\frac{256}{588\cdot90}
+\frac{113}{588\cdot120}
+\frac{12}{588\cdot90}
+\frac{2}{588\cdot36}
+\frac{4}{588\cdot6}\approx 0.017389
\]

\[
\boxed{\text{Expected accuracy}\;\approx\;1.739\%}
\]

If instead the guesser randomly chooses exactly two options each time (as this is the most common number of correct answers), the baseline expected performance on the benchmark would be 2.9\%. 

\[
\mathbb{E}[\text{accuracy}]
=\frac{256}{588}\cdot\frac{1}{15}
=\frac{64}{2205}
\approx 0.0290249
\]

\[
\boxed{\text{Expected accuracy}\;\approx\;2.902\%}
\]

\section{Effect of Filtering Hybrid Analysis Report Data on Malware Analysis Performance}
\label{sec:appendix_b2_context_size}

The majority of recent models included in our testing allowed for context sizes of up to 128,000 tokens. Because the Hybrid Analysis reports that form the basis of the questions in the Malware Analysis benchmark can be very large, some models may be unable to accommodate the full contents of these reports due to context window size limits. In order to make this benchmark available to shorter context models, we decided to experiment with filtering down the contents of Hybrid Analysis reports. 

In order to do so, we implemented an optional filtering parameter for this benchmark which restricts the fields from the Hybrid Analysis report that are retained prior to sending the report contents to the model under test, retaining only the most important keys and eliminating unnecessary information. This approach helps reduce data noise by retaining the most important information for the model under test to analyse. The important keys retained comprise:

\begin{lstlisting}
   important_keys = [
       "size",
       "type",
       "submit_name",
       "sha256",
       "av_detect",
       "vx_family",
       "threat_score",
       "threat_level",
       "verdict",
       "certificates_validation_message",
       "total_processes",
       "total_signatures",
       "file_metadata",
       "processes",
       "mitre_attcks",
       "network_mode",
       "signatures",
   ]
\end{lstlisting}

Eliminated keys such as uid, name, normalized, etc. are redundant and are already present elsewhere in the report. Within the key mitre\_attcks field, we only retain the following keys: tactic, technique, and attck\_id.

In addition to filtering out certain fields, we truncated descriptions in the signatures field to 50 characters and replaced all 32 char hashes with smaller unique letter hash values for better tokenization. Note that the truncation of description does lead to a loss of information in some cases where the description holds information on multiple executed commands.

\begin{figure}
    \centering
    \includegraphics[width=\textwidth]{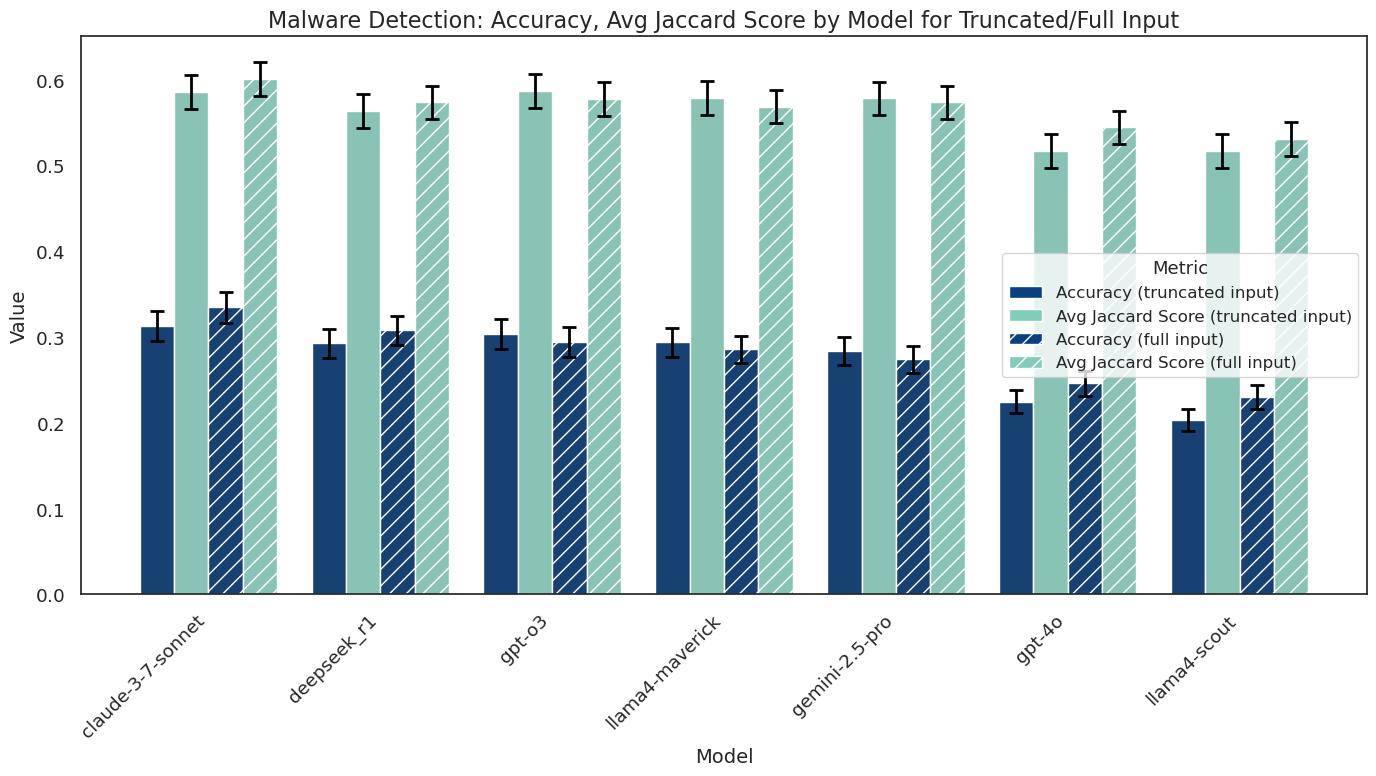}
    \caption{Per model evaluation scores for percent of multiple choice questions answered with perfect accuracy (blue) and average Jaccard index of model responses with the set of correct responses (teal) for the Malware Analysis benchmark. Solid bars reflect the performance of the model when provided only an abridged version of the Hybrid Analysis report, hatched bars show the performance when given the entirety of the output report. Error bars reflect 95\% confidence intervals computed using a normal approximation.
}
    \label{fig:b2-results-by-truncation}
\end{figure}

We present in Figure~\ref{fig:b2-results-by-truncation} some results of this benchmark evaluated on the same model with and without this filtering enabled, as a mini ablation study in order to understand how important it is for the model to have all of the information provided in the Hybrid Analysis report.

As shown in Figure~\ref{fig:b2-results-by-truncation}, we find that filtering the contents of the Hybrid Analysis report down to only ‘essential’ fields has a negligible impact on model performance, with most models performing a couple of percentage points better on topline metrics when given the full report, and in no case do we observe a statistically significant difference in performance.  This seems to indicate that models are primarily leveraging a subset of the available telemetry data for their reasoning, and implies that organizations which are resource or bandwidth constrained may still be able to reap some of the same benefits of using AI systems to process log data by implementing a prefiltering step.  On the other hand, the fact that models which allow for large volumes of inputs do not exhibit a decrease in performance relative to the case where they are fed only manually curated ‘essential’ components of log data seems to indicate that models are able to focus on the most relevant signals rather than becoming distracted or lost in the large amounts of noise—a promising sign for practical applications where it is preferable to automate processing of raw logs without investing resources to carefully design a prefiltering protocol.

Future work could explore which specific fields contribute most to model performance and whether this varies by attack type or security scenario, to help guide organizations that do wish to leverage AI systems on what the most relevant components of raw system logs to keep may be.

\section{Threat Intelligence Reasoning Additional Results}
\label{sec:appendix_b3_additional_results}

\begin{figure}
    \centering
    \includegraphics[width=\textwidth]{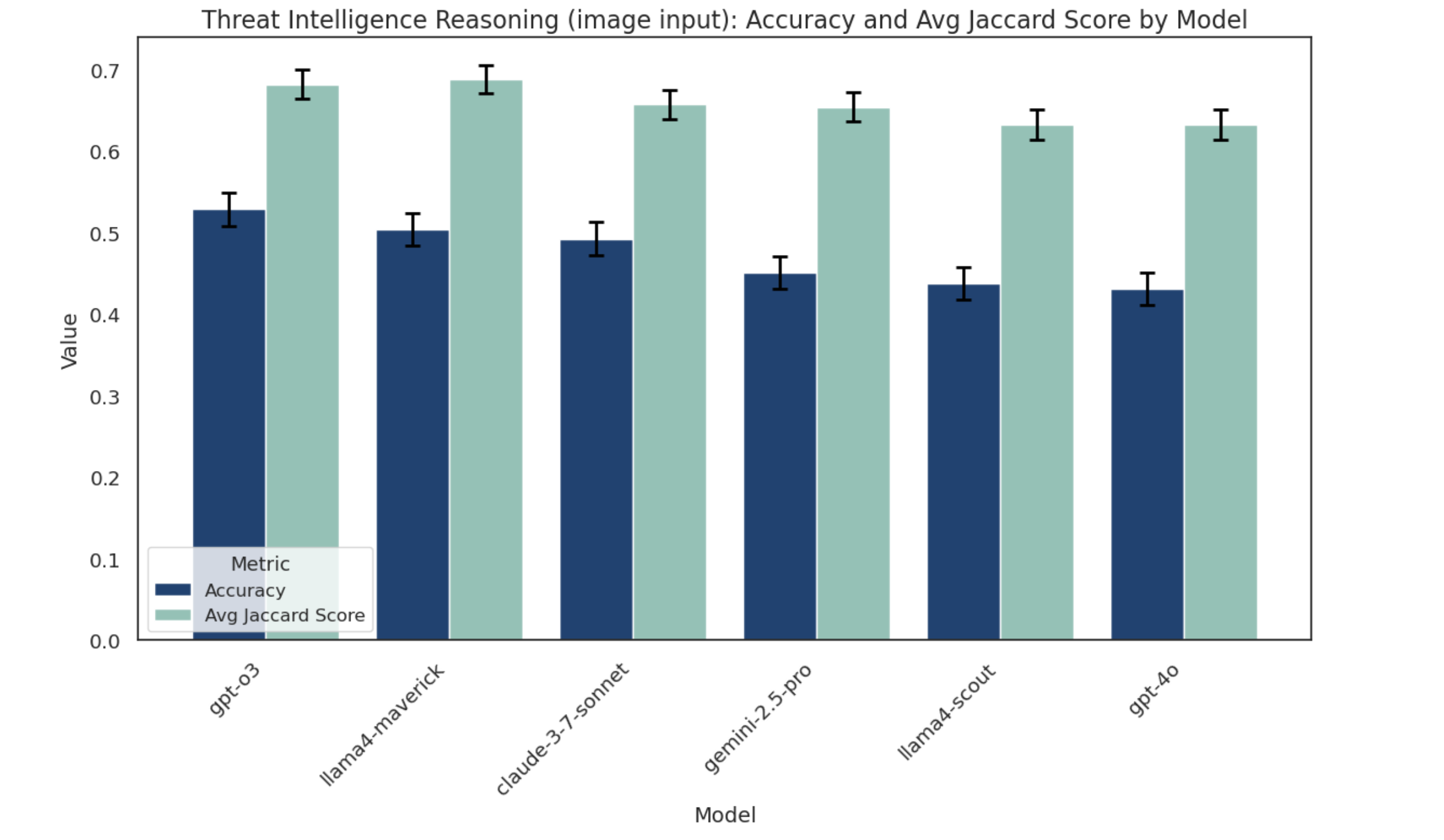}
    \caption{Per model evaluation scores for percent of multiple choice questions answered with perfect accuracy (blue) and average Jaccard index of model responses with the set of correct responses (teal) for the Threat Intelligence Reasoning benchmark. Error bars reflect 95\% confidence intervals computed using a normal approximation.}
    \label{fig:b3-topline-results}
\end{figure}

Figure~\ref{fig:b3-topline-results} shows per model evaluation scores for percent of multiple choice questions answered with perfect accuracy (blue) and average Jaccard index of model responses with the set of correct responses (teal) for the Threat Intelligence Reasoning benchmark. 

\section{Synthetic Data Generation Code for Malware Analysis Benchmark}
\label{sec:appendix_Python_b2}

\begin{figure}
    \centering
    \includegraphics[width=\textwidth]{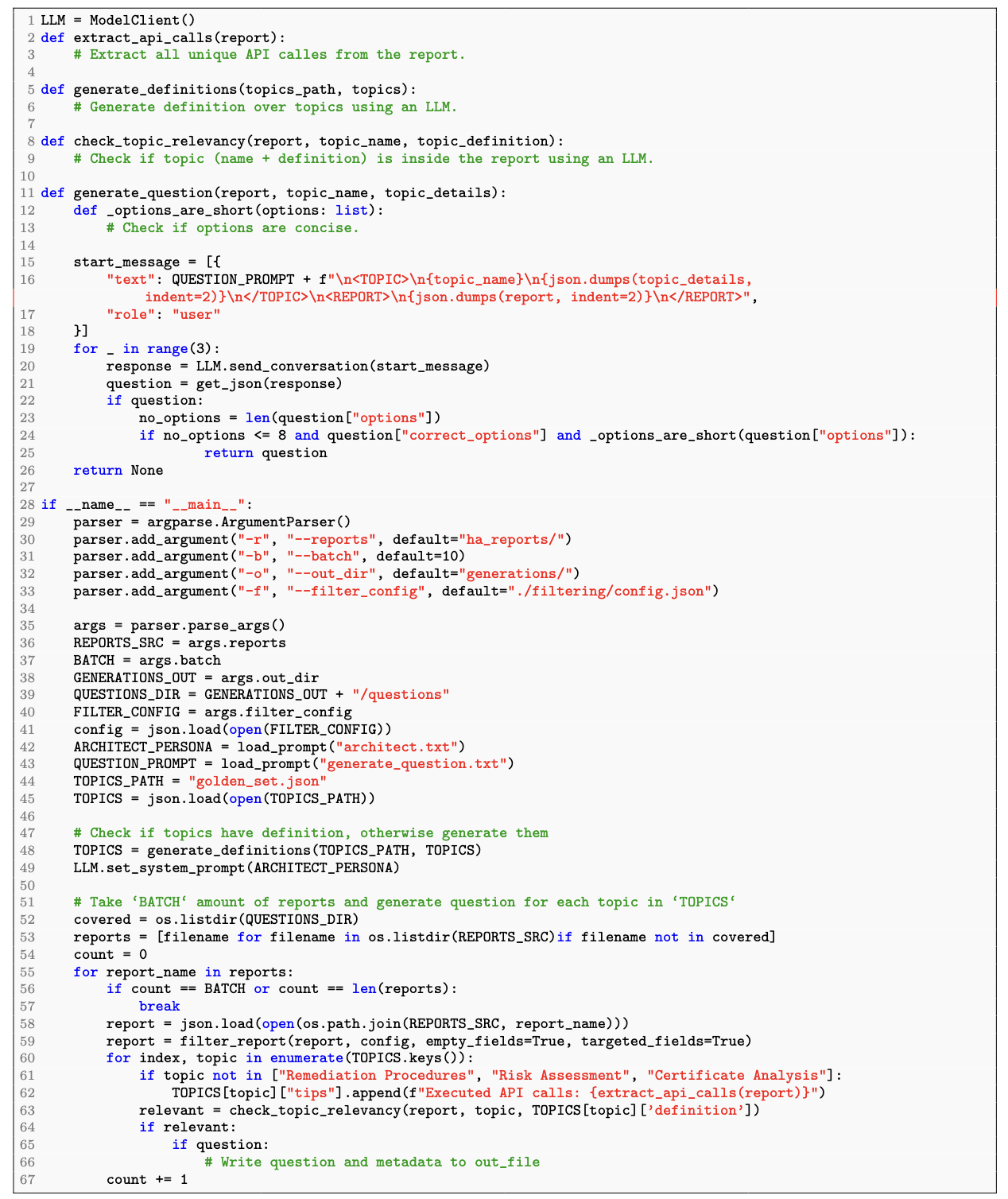}
    \caption{Python for malware analysis question creation.}
    \label{fig:b2-Python}
\end{figure}

\begin{figure}
    \centering
    \includegraphics[width=\textwidth]{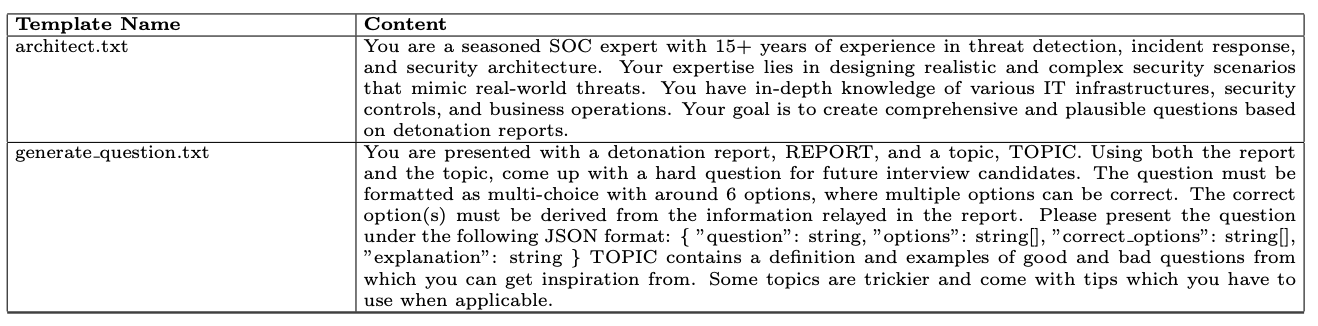}
    \caption{Template names for malware analysis benchmark.}
    \label{fig:b2-template-names}
\end{figure}

Figure~\ref{fig:b2-Python} shows Python for malware analysis benchmark question creation. Figure~\ref{fig:b2-template-names} shows names of templates. 

\section{Synthetic Data Generation Code for Threat Intelligence Reasoning Benchmark}
\label{sec:appendix_Python_b3}

\begin{figure}
    \centering
    \includegraphics[width=\textwidth]{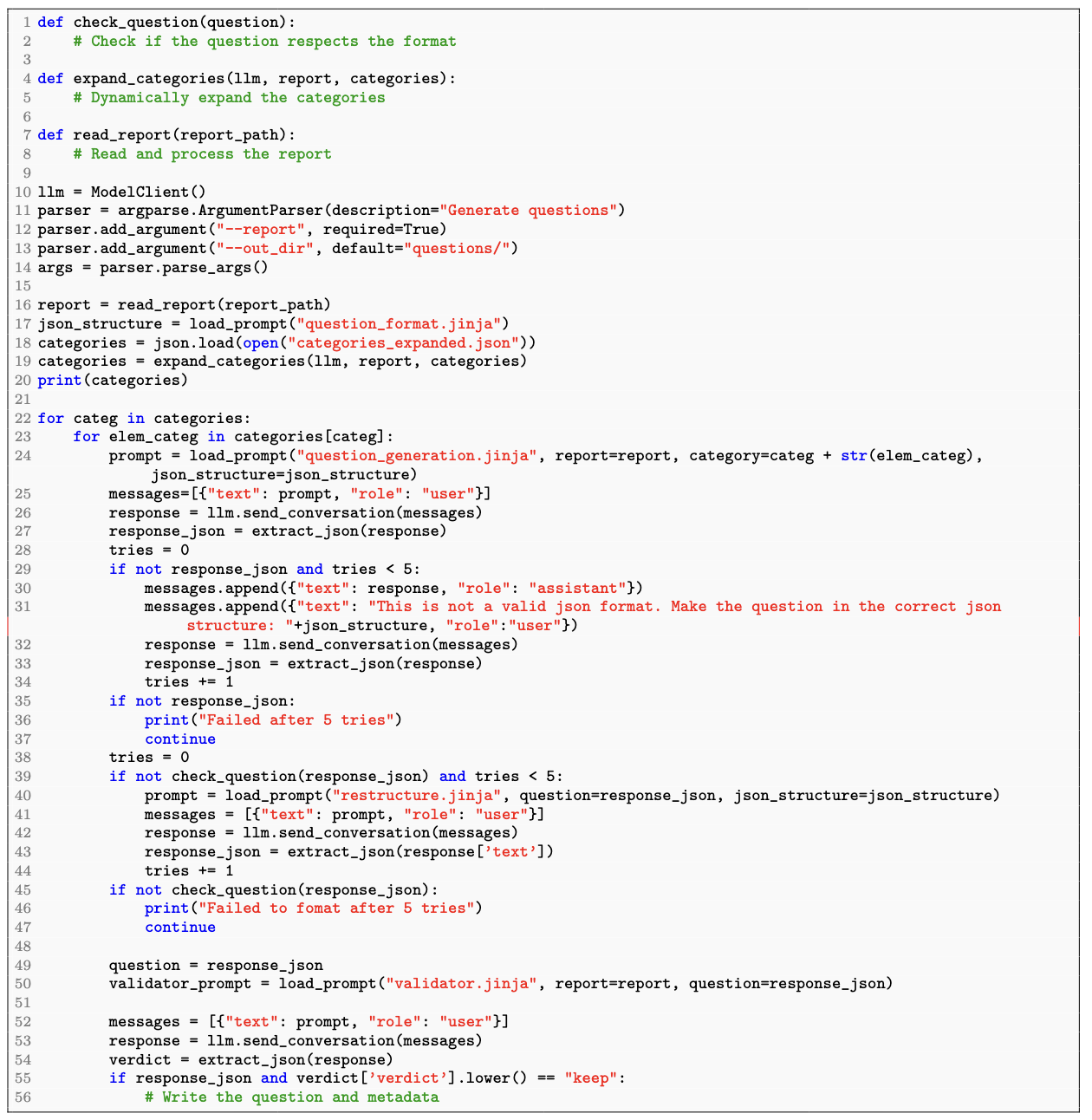}
    \caption{Python for category based generation. }
    \label{fig:b3-category-Python}
\end{figure}

\begin{figure}
    \centering
    \includegraphics[width=\textwidth]{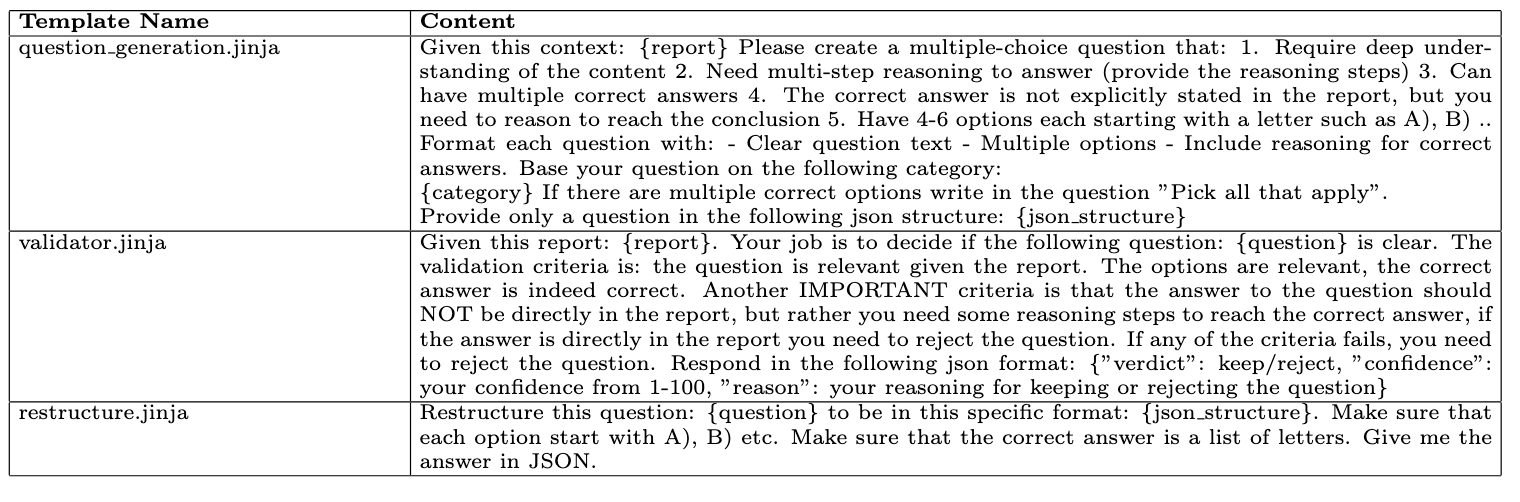}
    \caption{Templates for category based generation. }
    \label{fig:b3-category-template-names}
\end{figure}

\begin{figure}
    \centering
    \includegraphics[width=\textwidth]{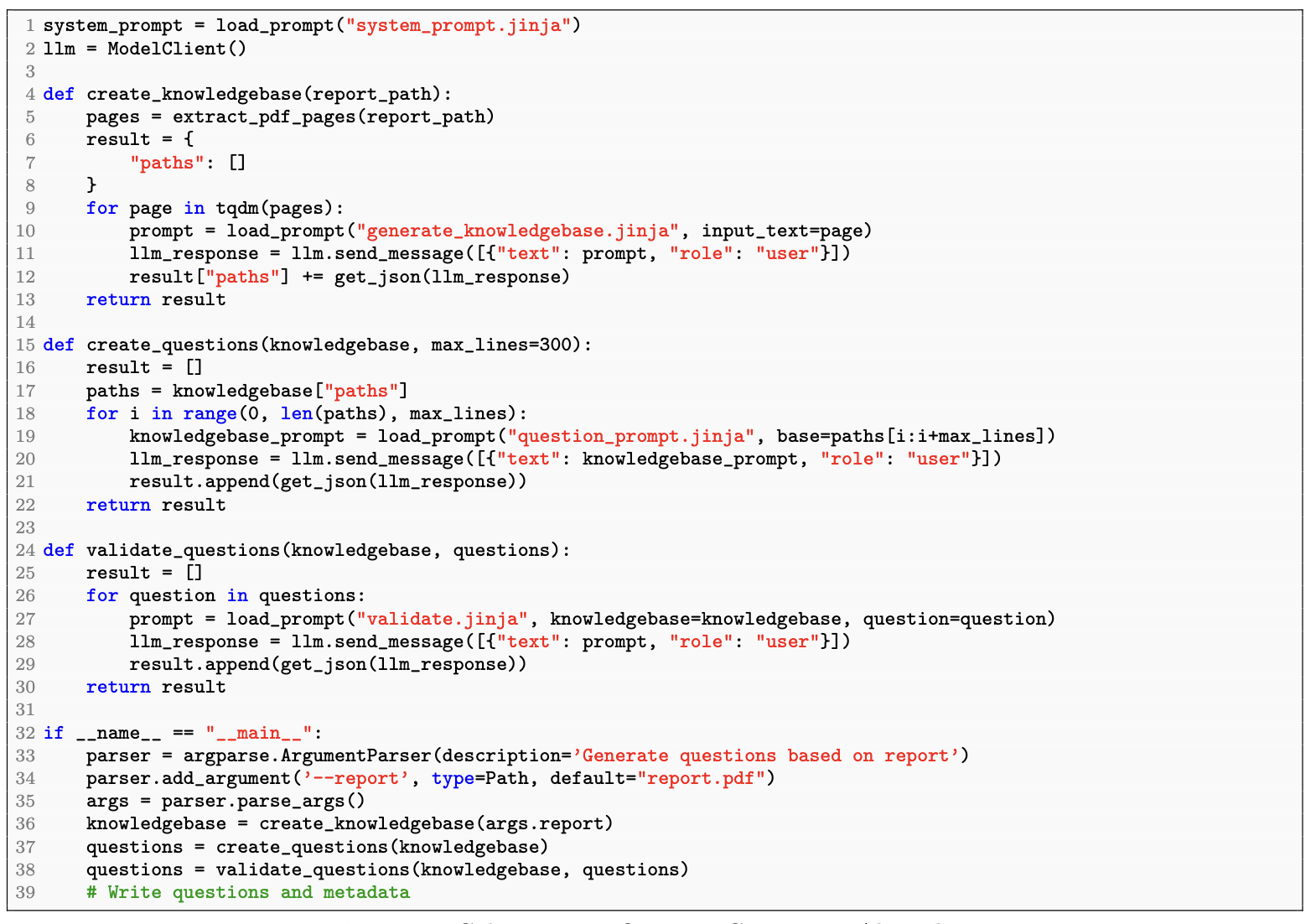}
    \caption{Python for relationship based generation.}
    \label{fig:b3-relationship-Python}
\end{figure}

\begin{figure}
    \centering
    \includegraphics[width=\textwidth]{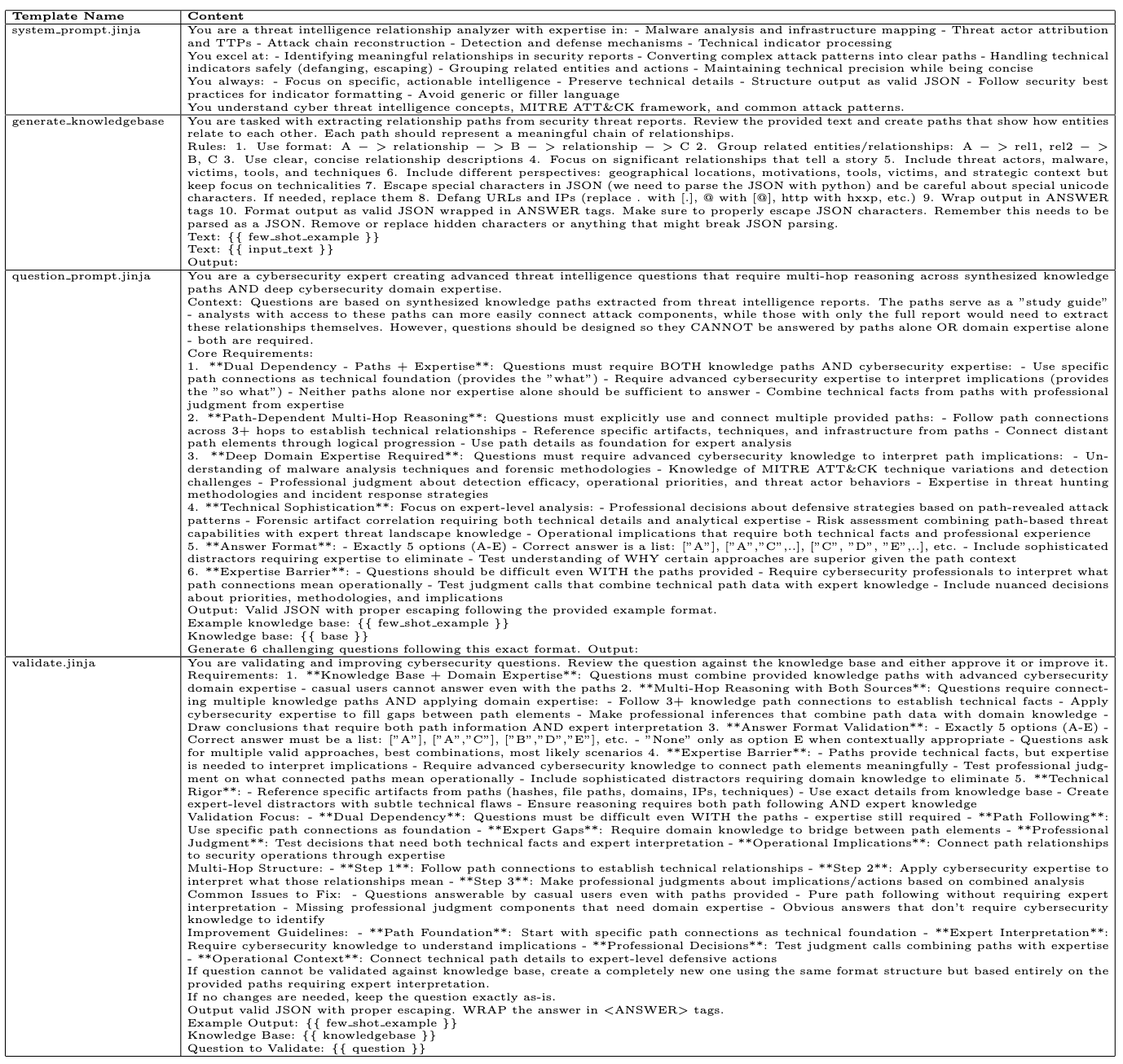}
    \caption{Template names for relationship based generation.}
    \label{fig:b3-relationship-template-names}
\end{figure}

Figure~\ref{fig:b3-category-Python} and Figure~\ref{fig:b3-relationship-Python} shows Python for category based and relatioship based generation. Figure~\ref{fig:b3-category-template-names} and Figure~\ref{fig:b3-relationship-template-names} show template names. 

